\documentclass[apj, letterpaper]{emulateapj}  

\pdfoutput=1

\usepackage{amsmath}

\usepackage{CJK}

\usepackage{xcolor}
\usepackage{color}

\begin{document}
\begin{CJK}{UTF8}{nsung}

\slugcomment{\bf}
\slugcomment{Submitted to the Astrophysical Journal}

\title{Migration and Growth of Protoplanetary Embryos II:
Emergence of Proto-Gas-Giants Cores versus Super Earths' Progenitor}



\author{ Beibei Liu\altaffilmark{1,2}, Xiaojia Zhang\altaffilmark{3}, 
Douglas N. C. Lin \altaffilmark{2, 3, 4}, and Sverre J. Aarseth$^5$}
\altaffiltext{1}{Department of Astronomy \& Astrophysics, Peking University, Beijing, 100871, China; bbliu1208@gmail.com}
\altaffiltext{2}{Kavli Institute for Astronomy \& Astrophysics, Peking University, Beijing, 100871, China}
\altaffiltext{3}{Department of Astronomy and Astrophysics, University of California, Santa Cruz, CA 95064, USA}
\altaffiltext{4}{Institute for Advanced Studies, Tsinghua University, Beijing, 100086, Beijing, China}
\altaffiltext{5}{Institute of Astronomy, Cambridge University, Cambridge, UK}

\begin{abstract}
\label{abstract}

Nearly $15-20\%$ of solar type stars contain one or more gas giant planet. 
According to the core-accretion scenario, the acquisition of their gaseous
envelope must be preceded by the formation of super-critical cores
with masses ten times or larger than that of the Earth.  It is natural 
to link the formation probability of gas giant planets with the supply 
of gas and solid in their natal disks.  However, a much richer population
of super Earths suggests that 1) there is no shortage of planetary 
building-block material, 2) gas giants' growth barrier is probably 
associated with whether they can merge into super-critical cores, and
3) super Earths are probably failed cores which did not attain sufficient
mass to initiate efficient accretion of gas before it is severely depleted.  
Here we construct a model based on the hypothesis that protoplanetary 
embryos migrated extensively before they were assembled into bona fide 
planets. We construct a Hermite-Embryo 
code based on a unified viscous-irradiation disk model and a prescription
for the embryo-disk tidal interaction.  This code is used to simulate
1) the convergent migration of embryos, and 2) their close encounters 
and coagulation. Around the progenitors of solar-type stars, the 
progenitor super-critical-mass cores of gas giant planets primarily 
form in protostellar disks with relatively high ($\gtrsim 10^{-7} 
M_\odot$ yr$^{-1}$) mass accretion rates whereas systems of super Earths
(failed cores)  are more likely to emerge out of natal disks with modest 
mass accretion rates, due to the mean motion resonance barrier and retention efficiency. 

\end{abstract}

\keywords{planetary systems -- planet--disk interactions -- methods: numerical }


\section{Introduction}
\label{Introduction}
 Over the past 2 decades, more than $1700$ exoplanets have been 
discovered and confirmed with radial velocity and transit 
surveys \citep{Schneider-et-2011,Wright-et-2011}. A widely 
accepted hypothesis for their origin is the sequential 
accretion scenario which assumes heavy elements in 
protostellar disks condense into grains, later coagulate 
into planetesimals and merge into protoplanetary embryos
\citep{Ida-Lin-2004a}.  Above a critical mass ($M_c 
\sim 10 M_\odot$), embryos become protoplanetary cores 
which accrete gas efficiently \citep{Pollack-et-1996}.  The 
fraction of stars containing gas giant planets, $\eta_J$,
is determined by whether this process can come to  
completion before gas is severely depleted (over a time
scale $\tau_{\rm dep} \sim$ 3-5 Myr in their natal 
disks).  After correcting for selection effects, existing
data suggest $\eta_J \sim 15-20\%$ around solar type 
stars \citep{Cumming-et-2008,Marcy-et-2008}.

In this paper, we discuss the formation probability of 
embryos with mass $M_p > M_c$.  The growth of their 
progenitor planetesimals evolves from runaway to 
oligarchic stage \citep{Kokubo-Ida-1998} and it is eventually 
impeded when they become dynamically segregated 
\citep{Lissauer-1987} with an isolation mass 
\begin{equation}
M_{\rm iso} 
\simeq (3/2) M_\ast ({ 4 k_0 M_d / 3 M_\ast} )^{3/2}
\label{eq:miso}
\end{equation}
where $M_d = \pi \Sigma_d r^2$ and $\Sigma_d$ are the 
characteristic mass and surface density of planetesimal
disk at a radius $r$, $M_\odot$ and $M_\oplus$ are the 
mass of the Sun and Earth respectively, $k_0 (\sim 10)$ 
is the normalized width of embryos' feeding zone $R_f = 
k_0 R_R$, $R_R \equiv a [2 M_p/(3M_\ast)]^{1/3} $ and 
$a$ are their Roche radius and semi major axis respectively.
In a minimum mass nebula (MMN) model \citep{Hayashi-1981, 
Ida-Lin-2004a}, the magnitude of $M_{\rm iso}$ is a
fraction of $M_c$ at the present location of Jupiter (5 AU).

We suggest that cores with $M_p > M_c$ were assembled 
in specific disk locations where $\Sigma_d$ acquired a local 
maximum value in excess of the power-law distribution in 
the MMN model. We assume that such concentration of 
building-block embryos was induced by the tidal interaction 
with natal disks which led to their type I migration
\citep{Goldreich-Tremaine-1979,Goldreich-Tremaine-1980}.  
It has been suggested that this process may lead to the accumulation
of building block material and enhance the growth of embryos
\citep{Nelson-2005,Lyra-et-2010, 
Paardekooper-et-2011, 
Horn-et-2012, Hellary-Nelson-2012,Pierens-et-2013}.

Based on the past simulations of embryo-disk interaction
(\cite{Paardekooper-et-2010,Paardekooper-et-2011}, hereafter PBK10, 
PBK11), we briefly recapitulate in \S2, the dependence 
of migration direction and speed on the gas surface density ($\Sigma_g$) and 
temperature ($T_g$) distribution. These simulations 
place a single embryo in a set of disk models with 
idealized power-law $\Sigma_g$ and $T_g$ distributions.  
With the publically available 2D hydrodynamic FARGO 
code \citep{Masset-2001}, \cite{Zhang-et-2014a,Zhang-et-2014b} 
(hereafter Z14a, Z14b) carried out simulations to
show that embryos do not significantly interfere each others' tidal
interaction with the disk and that type I migration 
indeed induces them to converge to some idealized trapping
radius $r_{\rm trap}$.

In this paper, we are interested in the possibility 
of collisions and coalescence of multiple embryos.  
We first use a self-consistent disk model 
(\cite{Garaud-Lin-2007}, hereafter GL07) to show 
the possibility that convergent migration leads 
them to a transition radius ($r_{\rm trans}$) between
the viscously heated inner and irradiated outer
regions of their natal disks. Since embryos' gravity 
does not significantly modify the disk structure, 
it is sufficient to assume the disk gas is in a 
hydrostatic and thermal equilibrium in the direction 
normal to the disk plane. In hydrodynamic simulations, 
any steady-state radial distribution of $\Sigma_g$ can be 
approximately maintained with an artificially specified 
viscosity prescription (Z14a). In subsequent studies, 
we will examine embryos' evolution during the depletion 
of the disk over a time scale $\tau_{\rm dep}$ of 
3-5 Myr when ${\dot M}_g$ declines by much larger 
magnitude than $L_\ast$ so that $r_{\rm trap}=r_{\rm trans}$ 
decreases from a few AU's to the stellar proximity 
(\cite{Kretke-Lin-2012}, hereafter KL12).  Such 
long-term changes in the boundary conditions will pose 
a challenge for a full scale multi-dimensional 
hydrodynamic simulation.  
  
In order to carry out the main technical study of this multi-
physics, multi-length-scales, and multi-time-scales problem, 
we construct a Hermite-Embryo scheme.  This code combines 
the calculation of multiple embryos' dynamical interaction 
\citep{Aarseth-2003} and the evaluation of their tidal interaction
with natal disks.  Based on the verification that embryo-disk
tidal torque is not affected by inference between multiple embryos 
(Z14a), we separately apply a torque prescription for each embryo 
(PBK10,11). In the calculation of the torque strength ($\Gamma$), 
we use a self-consistent model (GL07, KL12).

We verify, in \S3, that the Hermite-Embryo code reproduces 
individual embryos' type I migration rate ($\dot a$) and 
direction found with hydrodynamic simulations. We present
simulations with representative disk models.  In disks
with modest accretion rates (${\dot M}_g 
\lesssim 10^{-8} M_\odot$ yr$^{-1}$), embryos converge into convoys
of super Earth.  The migration time-scale ($\tau_a 
= a/{\dot a}$) is longer than the libration time ($\tau_{\rm lib}$)
of some lowest-order mean motion resonances (MMR's). They have 
a tendency to capture each 
others' MMR and form a convoy of resonant super Earths (Z14a).

In \S4, we show that in disks with high
accretion rates ${\dot M}_g (\gtrsim 10^{-7} M_\odot$ yr$^{-1}$), 
torque is sufficiently strong to induce
embryos to bypass the mean motion resonance barrier, cross
each other's orbits and merge into cores with $M_p > M_c$.
We estimate the critical value of ${\dot M}_g$
which separates these outcomes around solar type stars. For this 
objective, a set of 2D simulations are adequate. Finally 
in \S5, we summarize our results and discuss their implications.    

\section{Migration of planets' building blocks}
\label{ methods}

Our core formation scenario is based on the assumption that 
they are assembled near $r_{\rm trap}=r_{\rm trans}$ where their 
progenitor embryos congregated through convergent type I migration. 

\subsection{Availability of planet-building blocks}
For initial conditions, we assume the prior emergence of a population 
of Earth-size embryos through coagulation \citep{Kenyon-Bromley-2009, 
Dullemond-Dominik-2005,Garaud-et-2013}, gravitational instability 
\citep{Goldreich-Ward-1973, Weidenschilling-Cuzzi-1993,Youdin-Shu-2002, 
Garaud-Lin-2004}, streaming instability \citep{Youdin-Goodman-2005}, 
vortice trapping \citep{Johansen-Youdin-2007}, or pebble accretion 
\citep{Lambrechts-Johansen-2012}. Provided their growth 
can bypass several potential barriers such as hydrodynamic drag 
\citep{Adachi-et-1976} or collisional fragmentation 
\citep{Leinhardt-Richardson-2005,Stewart-Leinhardt-2009}, 
they may grow into dynamically segregated embryos with isolation 
mass $M_{\rm iso}$ shown in Eq. [\ref{eq:miso}].

Due to uncertainties in the opacity law, it is 
difficult to reliably extract from observations 
the value of $\Sigma_d$.  But, the gas accretion
rate ${\dot M}_g$ can be obtained from the UV 
veiling and spectroscopic data. The diffusion stability 
of accretion disks requires $\Sigma_g$ to be an
increasing function of ${\dot M}_g$ \citep{Pringle-1981}.
Any assumed $\Sigma_d - {\dot M}_g$ correlation would 
require an additional assumption: the dispersion in 
the metallicity of the disk gas ($Z_d$) is much 
smaller than that of ${\dot M}_g$.  It is tempting 
to extrapolate a dispersion in $\Sigma_d$ from 
the observed range of ${\dot M}_g (\sim 10^{-6}
-10^{-10} M_\odot$ yr$^{-1})$ among classical 
T Tauri stars \citep{Hartmann-et-1998} and to assume 
that gas giants are formed in disks with high 
$\Sigma_d$ and ${\dot M}_g$'s.

An alternative minimum planetary building block scenario
is to extrapolate a population of Earth mass embryos from 
the Kepler data \citep{Batalha-et-2013}.  After taking into
account its well understood observational selection effects
\citep{Dong-Zhu-2013}, transit search with this 
data set of controlled targets reveals the common existence 
of super Earth candidates with sizes in the range of $R_p \simeq 1-4 
R_\oplus$ where $R_\oplus$ is the Earth's radius. A vast majority of 
these multiple-planet systems are
most likely to be genuine planets \citep{Lissauer-et-2012}.  
Follow-up radial velocity (RV) surveys \citep{Marcy-et-2014} 
and mass determination  from the observed transit timing 
variations (TTV's) \citep{Wu-Lithwick-2013} have confirmed that: 
1) the fraction of stars ($\eta_E$) which harbor one or more 
super Earth candidates is much higher than $\eta_J$ 
\citep{Mayor-et-2011,Howard-et-2012,Fressin-2013}, and 2) in 
contrast to the observed correlation \citep{Fischer-Valenti-2005, 
Santos-et-2004,Johnson-2010} between $\eta_J$ and the 
metallicity of their host star ($Z_\ast$), the magnitude 
of $\eta_E$ appears to be independent of $Z_\ast$ and $M_\ast$ 
\citep{Wang-Fischer-2013,Buchhave-et-2012,Buchhave-et-2014}.

In the most up to date Kepler data release, there 
are 348 confirmed multiple-super Earth systems 
with $M_p < 100 M_\oplus$ and $N \geq 2$.  Among them, 
119 systems contain $N\ge3$ members.  Their confirmation 
is based on the upper mass limits obtained with the 
follow-up RV surveys or TTV measurements.  There are another
792 additional multiple Kepler objects of interests (KOI's) 
listed in the NASA Exoplanet Archive. All but 66 of these 
unconfirmed KOI's contain solely planetary candidates
with $R_p <10 R_\oplus$. Follow-up observations of these 
KOI's are needed to establish their planetary identity and to 
determine of their mean composition \citep{Wolfgang-Lopez-2014}.

\begin{figure}[htbp]
\includegraphics[width=1.\linewidth,clip=true]{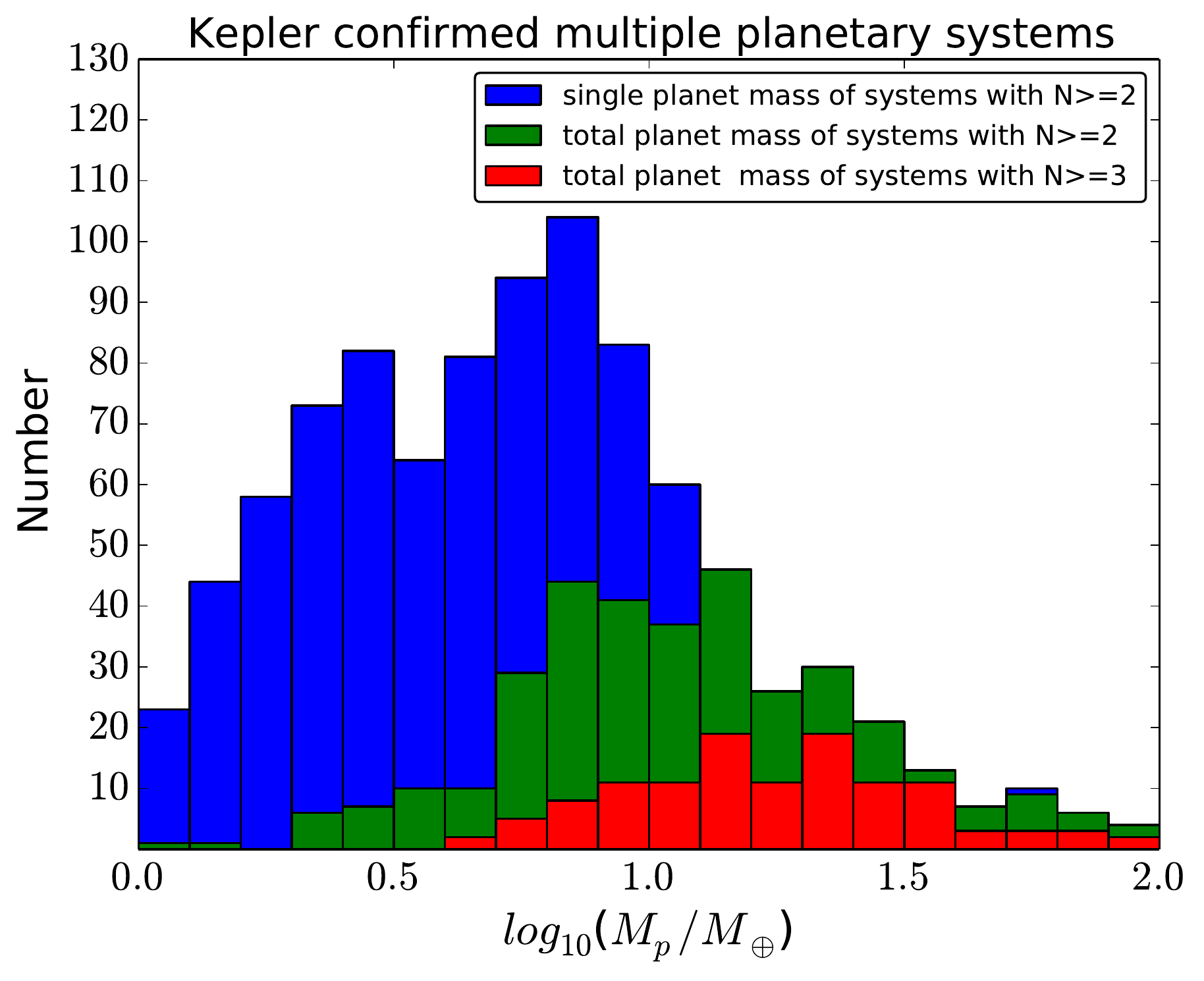}
\caption{
The planet mass distribution of multiple  exoplanetary systems in confirmed 
Kepler sample. Blue, green, red lines represent  single planet's mass 
($M_p$) in multiple systems with $N\geq 2$, total planetary system's 
mass ($M_s$) in multiple systems with $N\geq 2$ and with $N\geq3$ respectively.  
The masses are  extrapolated from 
$R_p$ with Eq. [\ref{eq:radiusmass}]).}
\label{fig:fig6}
\end{figure}

We plot the distribution of individual planets' $M_p$ and the 
total masses $M_s$ of multiple planetary systems for the 348 
confirmed Kepler multiple planetary systems (Fig. \ref{fig:fig6}).
Although the mass of some Kepler planets have been obtained 
with the RV or TTV measurements, most others do not have any
measured dynamical information.  We extrapolate their $M_p$ 
from a mass-radius relation 
 \begin{equation} 
R_{\rm p} \simeq (M_{\rm p}/M_{\oplus}) ^{1/2.06} R_{\oplus}.
\label{eq:radiusmass}
\end{equation}
which is empirically fitted to the solar system planets 
\citep{Lissauer-et-2011}.

Figure \ref{fig:fig6} shows that even though 
some individual super Earths may have extrapolated
density lower than that of the Earth and $M_p< M_c 
(\simeq 10M_{\oplus})$, the total mass $M_s$ of most
multiple KOI systems around individual host stars exceeds $M_c$. 
The total available building block materials in many multiple systems 
are enough to form the critical mass cores, but most of them do not 
evolve into gas giant planets. We interpret these data to imply that 
the lack of gas giants around most solar type stars may be due to the 
inability for sufficient fraction of all available building block 
materials to be collected into a few cores (with$ M_p \geq M_c$) 
rather than a lack of heavy elements in their natal disks. Based 
on this inference, we investigate the formation efficiency of cores.

In order to minimize the diverse statistical bias introduced 
by various survey methods, Figure \ref{fig:fig6} contains only
Kepler's confirmed multiple planets.  However, similar 
mass distributions are obtained with either all 792 unconfirmed 
KOI's or with all known multiple-planet systems including 
an additional 95 multiple planet systems which were discovered 
by RV or ground-based transit surveys.

\subsection{Embryos' type \uppercase\expandafter{\romannumeral1} Torque 
Formulas}
\label{sec:torque}


Embryos do not have sufficient mass to significantly perturb the disk 
structure \citep{Lin-Papaloizou-1993}.  But they can excite waves near
their Lindblad and corotation resonances.  Based on the assumption that
these waves are dissipated as they propagate through the disk, they 
lead to Lindblad and corotation torque which induces their migration.
Past simulations (PBK10, 11) showed that embryos' migration rate and 
direction depend on the $\Sigma_g$ and $T_g$ distributions. We first 
describe the analytic approximation of the embryo-disk torque.

Linear wave analysis indicate that there is an imbalance in the 
torque exerted by isolated embryos  at their Lindblad 
resonance on the disk region interior and exterior to their orbits
\citep{Tanaka-et-2002}. The net differential Lindblad torque generally 
causes embryos to undergo inward migration. Embryos also impose 
corotation torque on the disk gas which follows the horseshoe 
stream lines. With small speeds relative to the embryos, gas in 
this region interacts strongly with them. The sign of the corotation 
torque is determined by the local vortensity $(\zeta
=\Sigma_g/\Omega)$ of the gas.  The net (Lindblad plus corotation) 
residual torque determines the direction and speed of type I migration.  

In an inviscid disk, the initial local vortensity and entropy gradient
are erased along the horseshoe stream lines as gas is mixed within a
few libration periods \citep{Balmforth-Korycansky-2001}. This effect
leads to the saturation (weakening) of corotation torque. The angular
momentum transport induced by disks' turbulent Reynold's stress
leads to both angular momentum and entropy diffusion and the retention
of their intrinsic distribution.  The competition between planets' 
gravitational perturbation on the nearby stream lines and the mass
flow across the corotation region can be easily captured with an idealized 
linear analytic treatment.  

Extensive 2D numerical simulations (PBK10, 11) of tidal interaction 
between isolated planets in disks with intrinsic power-law 
surface density and temperature distributions ($\Sigma_g \propto 
r^s$  and $T_g \propto r^\beta$) have provided a data base 
to parameterize the net torque into the following form
\begin{equation}
\Gamma = f_\Gamma (s, \beta, p_\nu, p_\chi) \Gamma_0
\end{equation}
where the magnitude of the linear torque expression is
\begin{equation}
\Gamma_0 =(q/h)^2 \Sigma_p r_p^4 \Omega_p^2.
\label{eq:gamma0}
\end{equation}
Here $q= M_p/M_\ast$ is the mass ratio between the embryo and
its host star, and $h=H/r_p$ is the aspect ratio between the disk thickness ($H=
(R_g T/\mu)^{1/2} / \Omega_p$) and the location of the planet $r_p=a$.  

The maximum value and sign of the torque coefficient $f_\Gamma$ are 
determined by $s$ and $\beta$.  It includes the sum of all components of 
the Lindblad and corotation torque.  The full strength of the corotation
torque is determined by the $\zeta$ gradient which is is a function 
of $s$ and $\beta$ and the extent of its saturation is determined by 
the dimensionless parameters $p_\nu = (2/3) (R_e x_s^3) ^{1/2}$ 
and $p_\chi = (R_e x_s^3/P_t) ^{1/2}$ where 
$R_e \equiv \Omega_k r_p^2 / 3 \pi \nu$ and $P_t \equiv \nu/\xi$ 
are the Reynolds and Prandl numbers respectively, $\nu$ and $\xi$ are 
the viscous and radiative diffusion coefficients respectively, and 
$x_s \simeq (q/h)^{1/2}$ is the dimensionless width of the horseshoe 
regions.  Mass and entropy rapidly diffuse through the narrow horseshoe
region of low-mass embryos (with small $p_\nu$ and $p_\chi$) such that
their torque only affect the gas which u-turns very close to their azimuthal 
location.  In contrast, diffusion cannot cross the wide horseshoe region 
of high-mass embryos (large $p_\nu$ and $p_\chi$) within the 
libration time scale on the horseshoe orbit.  Consequently, the disk's
intrinsic vortensity gradient is effectively erased in this region and the
corotation torque is saturated. Maximum corotation torque is exerted by 
embryos with $p_\nu \sim 1$ or $p_\chi \sim 1$.  

Accretion disk models provide values for $s$, $\beta$, $\Sigma_g$, $T_g$, 
$h$, $\nu$,  $\xi$, $R_e$, and $P_t$ (see \S\ref{sec:disk}). 
Therefore we can evaluate  $x_s$, $p_\nu$, $p_\xi$, $f_\Gamma$, 
and total torque $\Gamma_{\rm tot}$ for embryos with 
mass $M_p$ at location $r_p (=a)$.  
The total torque leads to a net change in the semi major axis at rate
\begin{equation}
{d a \over d t} = {2 f_a q \over h^2} {\Sigma_g r_p^2 \over M_\ast}
r_p \Omega_k
\end{equation}
where   $\Gamma_{\rm tot} = f_a \Gamma_0$
and  $  f_a = \sum   f_{\Gamma,i}  $, the index $i$ refers to different
torque components including differential Lindblad ($f_{Lb}$), linear and nonlinear 
horseshoe and corotation torque.  In principle, the inner and outer Lindblad 
resonances of embryos located near $r_{trans}$ are located in regions of 
the disks with different values of $s$ and $\beta$.  But the separation 
between these resonances and the width of the corotation region are 
much smaller than both $a$ and the scale length over which
$s$ and $\beta$ changes (Z14b).  It is adequate to adopt a smoothing function
for the transition of $s$ and $\beta$ and use the value of $f_{\Gamma, i}$ at
$r_p=a$.  
We can also obtain the rate of change in eccentricity 
\begin{equation}
{1 \over e} {d e \over d t} = {2 f_e q \over h^4} {\Sigma_g r_p^2 \over M_\ast}
\Omega_k
\end{equation}
where we assume both inner and outer Lindblad torque lead to eccentricity 
damping so that 
$f_e = \sum \vert f_{\Gamma,i}  \vert $
 \citep{Goldreich-Tremaine-1980,Artymowicz-1993,Goldreich-Sari-2003}.    

\subsection{A Self-Consistent Model of Evolving Protostellar Disks}
\label{sec:disk}

The $\Sigma_g$ distribution is determined by the efficiency of angular 
momentum transfer \citep{Lynden-Bell-Pringle-1974,Ruden-Lin-1986}. The 
dominant angular momentum transfer mechanism in accretion disks is 
turbulent-induced viscous stress \citep{Shakura-Sunyaev-1973}.
The most likely cause for turbulence in inner ($< 0.1$ AU) regions 
of protostellar disks is magneto-rotational instability 
\citep{Balbus-Hawley-1991}.  Between $\sim 0.1-10$ AU, there is a dead 
zone where the ionization fraction is low and the field cannot diffuse 
through near the disk midplane. But, MHD turbulence is prevalent in 
the disk surface layer which is  ionized by the stellar 
irradiation and cosmic rays  \citep{Gammie-1996}. 

For our application, we need to carry out simulations over $10^{5-6}$
orbits.  It is computationally practical to construct a relatively 
simple disk model which robustly reproduces the generic outcomes 
of embryos' type I migration and apply it to the Hermite-Embryo code.
We adopt a self-consistent disk model of \cite{Garaud-Lin-2007}  
(hereafter GL07) based on the assumption that the inner region of 
the disk is heated by viscous dissipation whereas the outer region is 
heated by stellar luminosity ($L_\ast$). For these models, 
$r_{\rm trap}$ coincides with the transition radius
$r_{\rm trans}$ which separates these two regions (KL12). 

Based on the conventional $\alpha$ prescription for viscosity, we assume
\begin{equation}
\nu = \alpha_\nu C_s H = \alpha_\nu R_g T_g / \mu \Omega_k
\label{eq:nu}
\end{equation}
where $H$ is the scale height in the direction normal to the disk plane, 
$R_g$ and $\mu$ are gas constant and molecular weight. Typical
magnitude for the dimensions turbulent efficiency factor $\alpha_\nu \sim 
10^{-3}$.  In the layered regions, the efficiencies of angular momentum 
transfer, mass and entropy diffusion are miniscule at the midplane and 
modest near the surface of the disk. This height ($z$) 
dependent structure can be approximated by the standard value of 
$\alpha_\nu (\sim  10^{-3})$ for the surface layer ($\alpha_H 
\sim \alpha$ at $z \sim H$) and an order of magnitude smaller 
value ($\alpha_M \sim 0.1 \alpha$) for the midplane region beneath 
the partially ionized layer.  

For computational simplicity, the disk model we adopt in this paper 
is based on the assumption that it
evolves in a quasi steady state in which ${\dot M}_g (=3{\pi} \Sigma_g 
\nu) $ is independent of radius but declines exponentially over the gas 
depletion time scale such that 
\begin{equation}
{\dot M}_g = {\dot M}_0   e^{- t/\tau_{\rm dep}}
\label{eq:taudep}
\end{equation}
with $\tau_{\rm dep} \sim 3-5 $Myr. 

In \S\ref{sec:formation} and \S\ref{sec:merger}, we show that a critical 
criterion for the formation of super-critical cores is sufficiently high 
${\dot M}_g$ or $\Sigma_g$.  
Since ${\dot M}_g$ monotonically declines, the value of 
${\dot M}_g$ during the early epoch of embryo formation 
determines the outcome of the migration and merger process.  
For stable accretion disk models, the viscous diffusion time 
scale generally increases with the disk radius such that their 
inner regions generally establishes a quasi equilibrium state.
We are mostly interested in disk regions not much beyond 
$r_{\rm trap}$ (a few AU's), a quasi steady state would
be established if disks extend well beyond $\sim 10$ AU 
\citep{Birnstiel-Andrews-2014}.  
For simulations which indicate that the migration time scale 
is a fraction of $\tau_{\rm dep}$ (see \S \ref{sec:formation} 
and \S \ref{sec:merger}), we do not expect the disk structure to 
evolve significantly during embryos' migration.

Based on these prescriptions and an idealized opacity law in which 
$\kappa =\tilde{\kappa_0} T_g$, GL07 obtained $\Sigma=\Sigma_0 r_{AU}^s$
and $T_g = T_0 r_{AU}^\beta$, where $ r_{AU} =(r / 1{\rm AU})$, with  
\begin{equation}
\Sigma_0 = 240 \ \alpha_3^{-3/4} \kappa_0^{-1/4} m_\ast^{1/8} 
{\dot m}_9^{1/2} \ {\rm g} \ {\rm cm}^{-2},
\label{eq:sigma0vis}
\end{equation}
\begin{equation}
T_0 = 120 \ 
m_\ast^{3/8} {\dot m}_9^{1/2} \alpha_3^{-1/4} \kappa_0^{1/4} \ {\rm K},
\label{eq:t0vis}
\end{equation}
$s= -0.375$ and $\beta=-1.125$ for the viscously heated inner region. In the 
above expression, the normalized quantities, $\alpha_3 \equiv 
\alpha_\nu/10^{-3}$,
$m_\ast \equiv M_\ast/M_{\odot}$, ${\dot M}_9 \equiv {\dot M}_g / (10^{-9} 
M_\odot {\rm yr^{-1}}$) and $\kappa_0  \equiv  \tilde{\kappa_0}/0.02$.  The opacity $\kappa$ is in  units of $\rm {cm^{2}g^{-1} }$ and  likely
to be correlated with  disk metallicity ($Z_d$). The 
corresponding aspect ratio 
\begin{equation}
h=0.025 m_\ast^{-5/16} {\dot m}_9 ^{1/4} \alpha_3^{-1/8} \kappa_0^{1/8}
r_{AU}^{-1/16}
\label{eq:hvis}
\end{equation}
implies this region of the disk is self shadowed. 

In the limit of relatively large ${\dot M}_d ( > 10^{-8} M_\odot 
\rm yr^{-1}$), the transition from viscous dissipation to surface 
irradiation takes place in regions where the disk is opaque to either
incident stellar irradiation or reprocessed radiation (or both). 
Using a self consistent treatment of the $h$ distribution, GL07
show that this region is relatively confined.  Outside this transition
region, the disk becomes optically thin and $T_g$ there can be approximated
by the local equilibrium temperature 
\begin{equation}
 \frac{A_{s}}{2} \frac{ L_{\star}}{4 \pi r^2}= \sigma T_{e}^{4},  
\end{equation}
where $T_e$ is the disk's effective temperature, $\sigma$ is 
the radiation constant, $A_{s}$ is the grazing angle of the 
disk, which can be expressed as $A_{s} \propto r \ d (H/r) 
/d r $.  In this limit,
\begin{equation}
\Sigma_{0} = 95 \ m_\ast^{9/14} {\dot m}_9 l_\ast^{-2/7} \alpha_3^{-1}
\ {\rm g} \ {\rm cm}^{-2},
\label{eq:sigmaoirr}
\end{equation}
\begin{equation}
T_{0} = 300 \ l_\ast^{2/7} m_\ast^{-1/7} \ {\rm K}
\label{eq:t0irr}
\end{equation}
with $s=-15/14$, $\beta=-3/7$, and $l_\ast = L_\ast/L_\odot$.  There is
no explicit dependence on $\kappa_0$ for $\Sigma_0$ and $T_0$ in this region.
For computational simplicity, we neglect the opaque region 
and determine the transition( trapping) radius
\begin{equation}
r_{\rm trans} \simeq 0.26 m_\ast^{0.74} l_\ast^{-0.41} {\dot m}_9^{0.72} 
\alpha_3^{-0.36} \kappa_0^{0.36} {\rm AU}
\label{eq:trans1}
\end{equation}
by matching $T_g$ from the viscously heated and the optically thin
regions.  

In this model, all the structural parameters, including $s$, $\beta$,
$\Sigma_g$, $T_g$, $h$, and $\nu$, are functions of $M_\ast$, 
${\dot M}_g$, $r$, and $\alpha_\nu$.  In the dissipation-dominated 
inner disk region, we assume that turbulent mass and heat transport 
would yield a unit effective Prandl number (ie $P_t \simeq 1$).  This
approximation simplifies the evaluation of $p_\chi$.

\subsection{Embryos' Migration through Protostellar Disks}
\label{sec:migdisk}

We now combine the disk model with the torque formula, neglecting
any feedback on the disk structure (KL12).  This approximation is 
justified by previous hydrodynamic simulations especially for 
embryos with $R_R < H$ \citep{Lin-Papaloizou-1986,Paardekooper-et-2010, 
Zhang-et-2014b}.  

For very low-mass (with $p_\nu <<1$) and high-mass (with $p_\nu
> > 1$) embryos, corotation torque is highly saturated and only 
differential Lindblad torque in equation (6) contribute to the total torque such that
\begin{equation}
f_{Lb} =1.7 \beta - 0.1 s - 2.5
\end{equation}
is negative in both the inner and outer disk regions.  In this limit,
embryos would migrate inward until they reach the inner boundary of the 
disk.  

In the irradiated outer regions of the disk, embryos migrate inward 
even when the corotation torque operates at  full strength.  But 
for the viscously heated inner region, the fully unsaturated 
corotation and horseshoe torque is not only stronger than the 
Lindblad torque but also induces embryos to migrate outward.  However
the full strength of the corotation torque $\Gamma_{cr}$ 
can only be realized for embryos
with a range of masses (KL12).  In the viscously heated 
inner regions of the disk, the optimum mass for embryos' outward migration 
is obtained from the requirement $p_\nu \sim 1$ (or $p_\xi \sim 1$).
From the expression for $p_\nu$, $p_\xi$, equations (\ref{eq:nu}) 
and (\ref{eq:hvis}), we find
\begin{equation}
M_{\rm opt} \simeq  m_\ast^{13/48} {\dot m}_9 ^{7/12} \alpha_3^{3/8} 
\kappa_0^{7/24} r_{AU}^{-7/48} M_\oplus.
\label{eq:mopt}
\end{equation}

There is a tendency for embryos with $M_p \sim M_{\rm opt}$ to 
migrate and converge to $r_{\rm trans}$. The location of 
$r_{\rm trans}$ depends on both ${\dot M}_g$ and $L_\ast$ 
(see eq. [\ref{eq:trans1}]). In equation (\ref{eq:mopt})
$M_{\rm opt}$ depends on ${\dot M}_g$ and $M_\ast$ more sensitively 
than on $r$ (see further discussions in the next section). 
In the standard $\alpha$ disk model, the range of $M_p$ (around 
$M_{\rm opt}$) which can avoid saturation of corotation torque 
and enable outward migration \citep{Baruteau-Masset-2013} is
   \begin{equation}
 0.32q^{3/2} h^{-7/2} \leq \alpha_\nu \leq  0.16 q^{3/2} h^{-9/2}.
\label{eq:unsat}
\end{equation}
The ratio between the upper and lower mass limits ($
\sim (2 h)^{-2/3}$) is generally a few. 
In the derivation of the above mass range, the magnitude of $\alpha_\nu$
is assumed to be independent of the distance $z$ above the midplane.
However, the width of low-mass embryos' horseshoe region ($x_s r $)
is smaller than the thickness ($H_{\rm dead}$) of the dead zone 
beneath the disk's active surface layer. From equation (\ref{eq:unsat})
we find that a small $\alpha_M$ (appropriate for the disk midplane) 
would substantially reduce the lower limit in the mass range of embryos 
with unsaturated corotation torque (KL12). Although embryos with 
$M_p < M_\oplus$ may have a positive $f_a$, their $\tau_a > 
\tau_{\rm dep}$ due to the $M_p$ dependence in $\Gamma_0$ 
(Eq. [\ref{eq:gamma0}]).  

We now check for self consistency of our no feedback assumption.  
A necessary condition for embryos to induce sufficiently strong 
perturbation is to open a gap \citep{Lin-Papaloizou-1986a} is $R_R > H$.  
For optimum-mass embryos, 
\begin{equation}
{ R_R (M_{\rm opt}) \over H} \sim 0.5
m_\ast^{5/72} {\dot m}_9 ^{-1/18} \alpha_3^{1/4} \kappa_0^{-1/36} 
r_{AU}^{1/72}
\end{equation}
such that gap formation may be marginally avoided as numerical 
simulations have shown.

\begin{table*}[!ht]
\centering
\caption{The properties of adopted models}
\begin{tabular}{|c|c|c|c|c|}
\hline
\hline
Model  &  Accretion rate ${\dot M}$ ($M_\odot$ yr$^{-1}$)  &  viscous $\alpha$ & include dead zone  \\
\hline
 A   &  $7\times 10^{-9}$  & $10^{-3}$  & NO \\
 B   &  $10^{-7}$  & $10^{-3}$  & NO \\
 C   &  $10^{-7}$  & $10^{-3}$  & YES \\
\hline
\end{tabular}

\begin{tabular}{|c|c|c|c|}
\hline
 Model & number of planets ($N_p$) & planet Mass  ($M_{\oplus}$)   & include planet-disk interaction  \\
\hline
A1 &  15   &  2.0 & YES  \\
A2 &  4   &  10.0  & YES\\
B1 &  2 (inner) +5 (outer)  &  5.0 (inner)+4.0 (outer)  & YES \\
B2 &  2 (inner) +5 (outer)   &  5.0 (inner)+4.0 (outer)  & NO \\
C1 &  15   &  2  & YES \\
\hline
\end{tabular} 
\label{tab1}
\end {table*}

\subsection{The Hermite-Embryo numerical scheme}

The torque prescription constructed by PBK10 is for single 
power-law $\Sigma_g$ and $T_g$ distribution.  In our disk model, 
the values of $s$ and $\beta$ change across $r_{\rm trans}$.  
Hydrodynamic simulations show that the strength and sign of 
the torque are not significantly modified by this more complex 
disk structure (Z14b).

The torque prescription is an approximation of the tidal interaction
between isolated embryos and their natal disks.  As the embryos 
converge, their horseshoe regions overlap.  The perturbation by
neighboring embryos may modify the gas stream lines and the saturation
condition.  Detailed 2D hydrodynamic simulations show that such
interference does not significantly modify the corotation torque and the 
prescription derived for isolated embryos continues to provide adequate 
approximation for a system of converging embryos (Z14b).  

Based on these justifications, we modify an N-body HERMIT4 code
\citep{Aarseth-2003} and construct a Hermite-Embryo code to 
include the effect of embryos-disk interaction.  
Gravitational interaction between representative embryos 
is calculated with a time-symmetric scheme of \cite{Kokubo-Ida-1998} 
and the Burdet-Heggie regularization (discussed in Aarseth 2003) is applied for the treatment of 
close encounters.  These features enable efficient and reliable 
integration of embryos dynamics on time scales comparable to 
$\tau_{\rm dep}$ (a few Myr).

Separate disk torque on individual embryos are added to the equation 
of motion such that 
\begin{equation}
\frac{\mathrm{d}  v_{\mathrm{\theta}}   }{\mathrm{d}t}  
=  \frac{ \Gamma_\mathrm{tot} }{M_{\mathrm{p} }r },
\end{equation}
\begin{equation}
\frac{\mathrm{d}  v_{\mathrm{r}}   }{\mathrm{d}t}  
= - \frac{   v_{\mathrm{r}}  }{ \tau_{\mathrm{e}}}.
 \end{equation}
The migration timescale is then given by 
\begin{equation}
\tau_{\mathrm{a}}  \simeq \frac{a}{\dot{a}} =
M_{\mathrm{p}} \sqrt { (G M_\ast a ) }   
/(2 f_a \Gamma_0).
\label{eq:tauar}
\end{equation}
where  a negative value  for $f_a$ 
implies orbital decay.  The timescale for eccentricity damping 
timescale is a factor of $\rm (1/h)^2 $ shorter \citep{Tanaka-et-2002,Kley-Nelson-2012} 
than the orbital decay timescale:
 \begin{equation}
 \tau_{\mathrm{e}}  \simeq \frac{e}{\dot{e}} 
  =h^2  M_{\mathrm{p}} \sqrt { (G M_\ast a ) }   
/(2 f_e \Gamma_0),
\label{eq:tauer}
\end{equation}
where $f_e$ is the coefficient for total Lindblad and 
corotation torque, 
$v_{\mathrm{\theta}}$ and $v_{\mathrm{r}}$ are the 
velocity in azimuthal and radial direction. 
 
The Hermite-Embryo code is well suited to simulate embryos' 
long-term interaction with each other and their natal disks. It can 
reproduce MMR capture and treat close encounters between
multiple embryos. For comparison with previous results in Z14a and Z14b, 
all the models presented in this paper are simulated in the 2D limit.
This approximation does not affect the condition for MMR capture. 
But it does reduce the collision frequency. During episodes when two 
embryos' separation becomes smaller than the sum of their physical 
radii ($R_p$ obtained from equation (\ref{eq:radiusmass}) ), we assume 
they merge with conservation of total mass and angular momentum.
In the 3D limit, the collision time scale is given by equation  (\ref{eq:radiusmass}).
But in the 2D (mono-layer) approximation 
$\tau_c$ is smaller by the reduction factor $f_{2/3D}$.
Currently, we neglect a small amount of angular momentum transfer 
between the embryos' spin and orbit.  This effect will be examined 
in a future follow-up study.

\begin{figure*}[htbp]
\includegraphics[width=0.5\linewidth,clip=true]{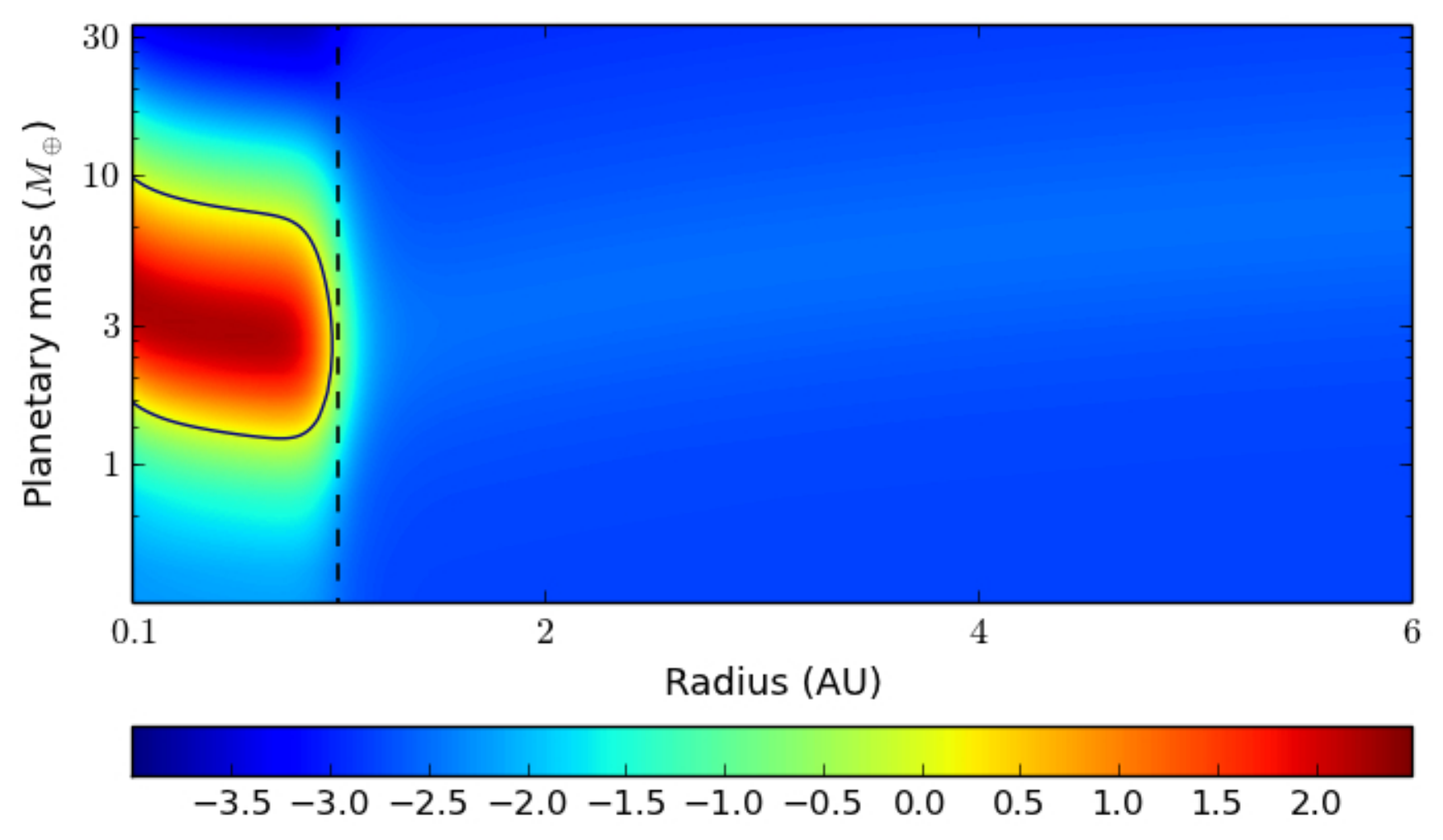}
\includegraphics[width=0.5\linewidth,clip=true]{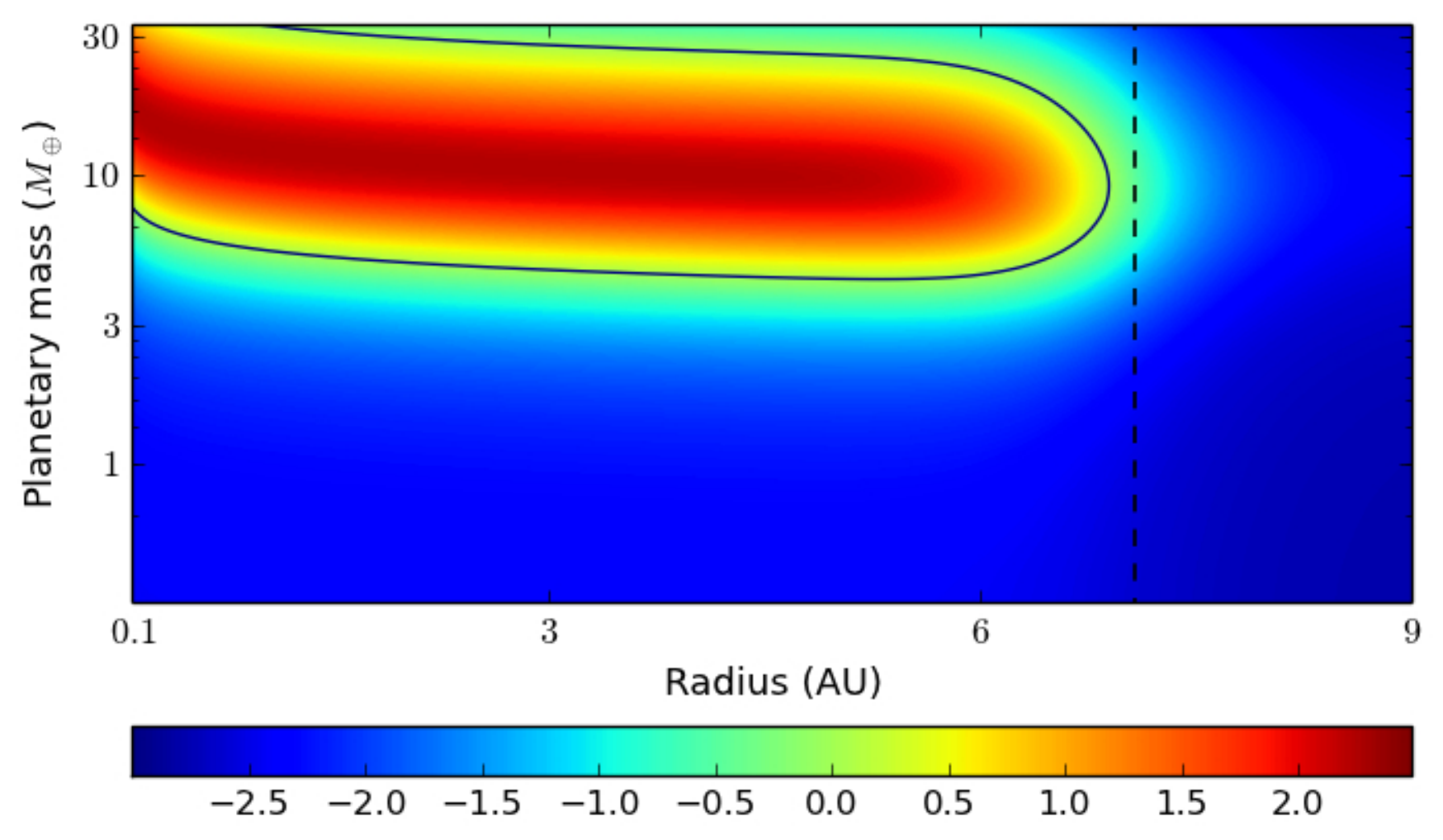}
\caption{
The type \uppercase\expandafter{\romannumeral1} 
migration coefficient ($f_a$) for a range of embryos'
mass at different location of the disk. Black dashed line represent the transition radius $r_{\rm trans}$.
{\bf Left}: model A with $\rm \dot{M}= 7\times10^{-9} 
\rm M_{ \odot}/\rm yr $, $\alpha_\nu =10^{-3}$, $ \Sigma_0=635 
\rm g/\rm cm^2$ and $r_{\rm trans}=1.05\rm AU$.
{\bf Right}: model B with $\rm \dot{M}= 1\times10^{-7} 
\rm M_{ \odot}/\rm yr $, $ \alpha_\nu =10^{-3}$, $ \Sigma_0=2400 \rm g/\rm 
cm^2$ and $r_{\rm trans}=7.10 \rm AU$.
}
\label{fig:fig1}
\end{figure*}

\section{Emergence of Resonant Super Earths}
\label{sec:formation}

\subsection{Preferential Destiny of Migrating Embryos}

Applying the disk model (GL07) into the torque prescription 
(PBK10, 11), we determine the coefficients $f_\Gamma$, $f_a$ 
and $f_e$ of the total type \uppercase\expandafter{\romannumeral1} 
torque, migration, and circularization rates.  Two panels 
in Figure~\ref{fig:fig1} show the radial distribution of $f_a$ 
for different mass embryos.  The black dashed line denotes 
$r_{\rm trans}$.  In the irradiated regions exterior 
to the black line, $f_a < 0$ and embryos of all masses 
migrate inward.  In the viscously heated inner region 
(interior to the black line), corotation torque, at 
its full strength, dominates the differential Lindblad 
torque and induces embryos to migrate outward.  However, 
we can use the PBK10 prescription to show that only within 
a limited range of $M_p$ around $M_{\rm opt}$
(Eq. [\ref{eq:unsat}]), embryos migrate outward because 
their corotation torque is not severely saturated.

In Figure~\ref{fig:fig1}, the disk parameter for 
model A is ${\dot M}_g = 7 \times10^{-9} M_\odot {\rm yr}^{-1}$,
 whereas for model B it is ${\dot M}_g = 10^{-7} M_\odot {\rm yr}^{-1}$. 
   In both models, $\alpha_\nu = 10^{-3}$ and $M_\ast = 1 M_\odot$. 
 These parameters are chosen to respectively
represent the advanced and active phases of disk evolution.
 For model A (left panel), embryos interior to 
$r_{\rm trans} =1.05$AU and with $M_p \sim 2-9 \rm M_{\oplus}$
migrate outward.  For model B (right panel), $r_{\rm trans} 
= 7.10$ AU.  Interior to $r_{\rm trans}$, embryos with mass in 
the red region (i.e. $M_p \sim 5-25 \rm M_{\oplus}$) migrate 
outward.  Both the optimum mass and mass range are in good 
agreement with the values estimated with equations (\ref{eq:mopt})
and (\ref{eq:unsat}). In a minimum mass nebula, $M_{\rm iso}$ is 
around a few $M_\oplus$ which falls within the outwardly migrating 
range in Figure \ref{fig:fig1} and these embryos have a tendency 
to migrate to and accumulate near $r_{\rm trans}$.

\subsection{Convoys of super Earths trapped in MMR}

\begin{figure*}[htbp]
\includegraphics[width=0.5\linewidth,clip=true]{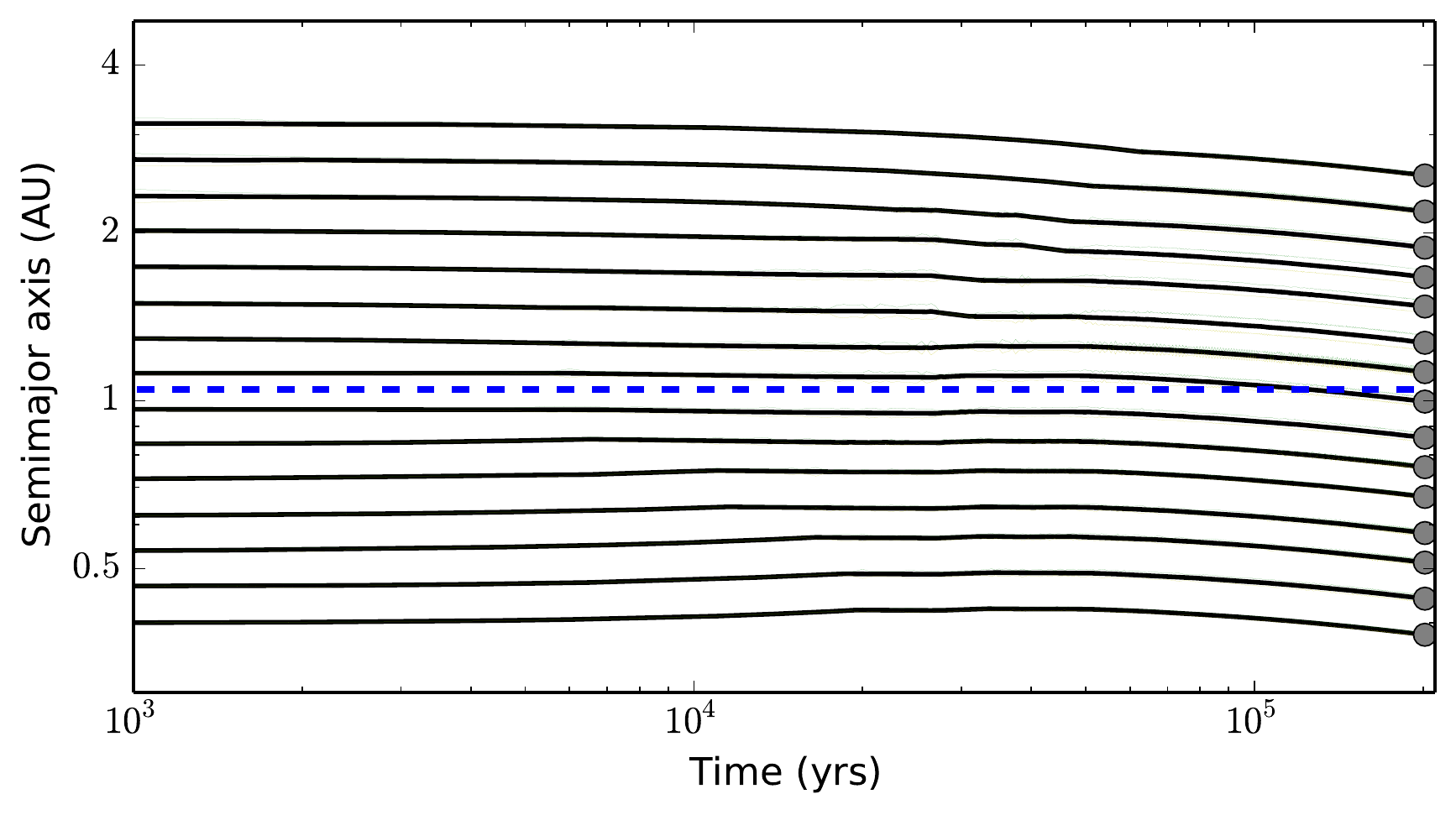}
\includegraphics[width=0.5\linewidth,clip=true]{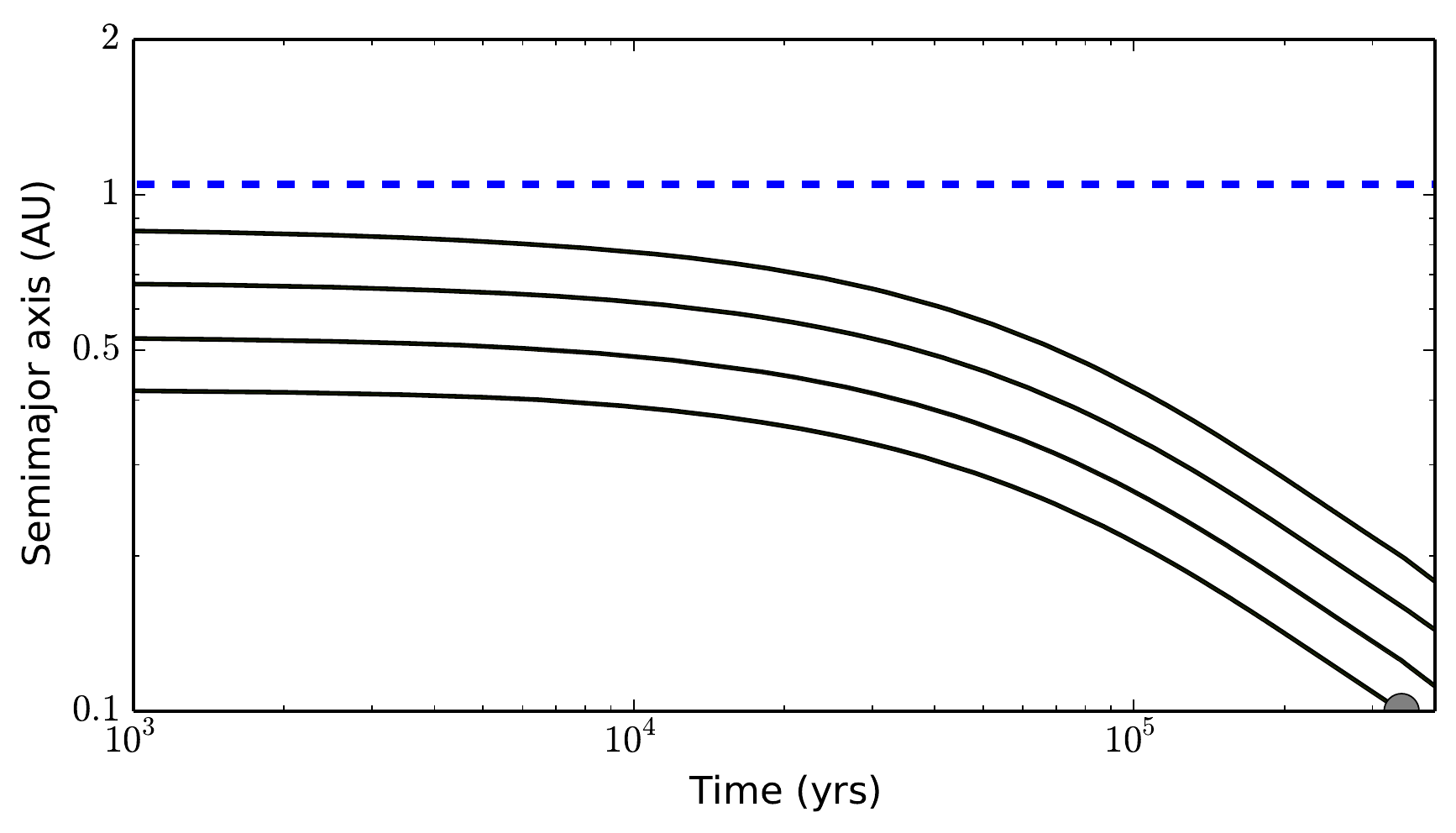}
\caption{
The mutual interaction between embryos and their natal disks. 
The black lines trace the evolution of embryos' semi-major axis and  
blue dashed line indicates the location of  $r_{\rm trap}$.  Green and 
yellow lines are embryos' apocentre and pericentre distance distance. 
Disk parameters are chosen to be those of model A. 
{\bf Left} model A1: A system of fifteen $2 M_{\rm \oplus}$ 
embryos which are initially separated by $10 R_{\rm R}$ between $0.6-3.7$AU. 
{\bf Right} model A2: A system of four $10 M_{\rm \oplus}$ embryos 
which are initially separated by $10 R_{\rm R}$ between $0.43-0.83$AU. 
}
\label{fig:fig2}
\end{figure*}

In order to examine embryos' concurrent interaction with 
each other and with their natal disk, we use the Hermite-Embryos 
code to compute the dynamical evolution of multiple embryos.  
Based on the boundary conditions in models A and B, we 
present the results of two series of simulations.

In model A1, we adopt 15 equal-mass ($M_p = 2 M_\oplus$) embryos. 
They are initially separated by $k_0=10$ with $a$ in the range of 
$0.4-3.1$AU. During the oligarchic growth, the embryos' isolation mass 
\begin{equation}
M_{\rm iso}\simeq 0.16
\left( {\Sigma_d \over 10 {\rm g 
\ cm^{-2}}} \right)^{3/2} \left( {r \over  {\rm 1 AU}} \right)^3 
\left( {M_\ast \over  1 M_\odot} \right)^{-1/2} M_\oplus.
\end{equation} 
Assuming a uniform normalized metallicity $Z_d$ in the disk
(with respect to the solar composition) for the viscously 
heated inner region, we find from equation (\ref{eq:sigma0vis}),
\begin{equation}
M_{\rm iso} \simeq 0.16 Z_d^{3/2} \kappa_0^{-3/8}
\alpha_3^{-9/8} m_\ast^{-5/16} 
{\dot m}^{3/4}_{-9} r_{AU}^{39/16} M_\oplus.
\label{eq:misoz}
\end{equation}
Note that $\kappa_0$ is a function of metallicity in micron 
grains whereas $Z_d$ is metallicity in condensed heavy elements.

The initial values of $M_p$ in model A1 are self consistent with
the disk parameters at the chosen location. 
All the embryos are assumed to be coplanar 
and their initial eccentricities follow a Rayleigh distribution  
 \begin{equation}
p(e) = \frac{e}{\sigma_0^2} \exp \left(  - \frac {e^{2}}{2\sigma_0^2} 
\right), \ \sigma_0= 0.02.
 \end{equation}
 Other orbital elements (argument of periastron, longitude of ascending node and mean anomaly)
 are  chosen randomly from $0^{\circ}- 360^{\circ} $.

The orbital evolution of these systems are all computed less than $5 \times 
10^{5}$ yrs.  Since this is shorter than the disk depletion time scale,
we adopt a steady state disk model.  
But, in all our models, the computed time is adequately long to 
simulate the embryos' migration and their potential collisions.  

Embryos' migration of Model A1 is plotted in the left panel 
of Figure~\ref{fig:fig2}.  They undergo convergent migration 
on a time scale of $\sim 6-8 \times 10^4$ yr as expected.  
However, their migration is stalled when they capture each 
other into their mutual MMR's. Although the total mass of 
the embryos ($30 M_{\rm \oplus}$) exceeds $M_c (\sim 10 
M_\oplus$) required for the onset of gas accretion, they 
form a compact convoy of super Earths with non intersecting 
orbits. The grey dots in  Figure~\ref{fig:fig2} are plotted at the endpoint evolution of each embryos, which is proportional to the mass of embryos.   

\subsection{Embryos' MMR Capture Condition}

The theory of MMR capture has been extensively developed by
\cite{Peale-1976} and \cite{ Murray-Dermott-1999}  (hereafter MD). 
A necessary condition for MMR capture is that the time 
scale $\tau_{\Delta a} (\sim \Delta a_{\rm res}/ {\dot a})$ 
for migration through their characteristic width ($\Delta 
a_{\rm res}$) is longer than their libration time scale 
$\tau_{\rm lib} \sim ( q f_{\rm res} e_{\rm res})^{-1/2}n^{-1}$, 
where $n$ is orbital mean motion and $f_{\rm res} = 
\alpha_a f_d( \alpha_a) $ (where $f_d$ is a function 
of the semi major axis ratio $\alpha_a$) from  Eq. 8.47 
in MD.  Within $\Delta a_{\rm res}$, an equilibrium 
eccentricity $(e_{\rm res})$ is maintained by a balance 
between its excitation during embryos' resonant migration 
\begin{equation}
e {\dot e}_{\rm exc} \sim {\dot a}/ a \sim 1/\tau_a
\end{equation}
(see Eq. 8.37 in MD) and its damping due the embryo-disk torque 
${\dot e}_{\rm damp} \sim e/\tau_e$ such that $e_{\rm res} \sim 
(\tau_e/\tau_a)^{1/2} \sim h$ (see Eqs. [\ref{eq:tauar}] and
[\ref{eq:tauer}]). From equation (8.58) in MD, we deduce
\begin{equation}
\Delta a_{\rm res} \sim ( qf_{\rm res} e_{\rm res} )^{1/2} a \sim ( qf_{\rm res} h )^{1/2}  a
\end{equation}
\begin{equation}
\tau_{\rm lib} \sim ( a / \Delta a_{\rm res} )  P / 2 \pi.
\end{equation}
With these dependencies, MMR's capture condition $\tau_{\Delta a}> 
\tau_{\rm lib}$ is reduced to 
\begin{equation}
\Sigma_g r^2 < \Sigma_{\rm res} r^2 \simeq f_{\rm res} h^3 M_\ast, 
\label{eq:sigmares}
\end{equation}
which is independent of embryos' mass.  In the above expression, 
$\Sigma_{\rm res}$ is defined to be the critical surface density
for resonant capture.  

The magnitude of $f_{\rm res}$ is the order of a few and it 
decreases with $\alpha_a$ (i.e. $f_{\rm res}$ is smaller 
for 3:2 than 2:1 resonance) such that it is possible for 
two embryos to enter into their 3:2 MMR even though they 
have failed to do so at their 2:1 MMR (Z14a).  In the proximity of 
$r_{\rm trans}$, we find (from eqs. [\ref{eq:sigma0vis}], 
[\ref{eq:trans1}] and [\ref{eq:sigmares}] ) that, 

\begin{equation}
\Sigma_g (r_{\rm trans}) r_{\rm trans}^2 \simeq
{\dot m}_9 ^{1.67} 
m_\ast^{1.33} l_\ast^{-0.67} \alpha_3^{-1.34} \kappa_0^{0.34} M_\oplus.
\label{eq:sigmartrans}
\end{equation}
with the critical condition for resonant trapping to be
\begin{equation}
{\dot m}_9 \simeq 6 f_{\rm res}^{0.95} m_\ast^{-1.33} \alpha_3^{0.97} 
\kappa_0^{-0.026} l_\ast^{0.70}.
\label{eq:modtrtrans}
\end{equation}



This analytic approximation confirms that during their convergent 
migration, embryos embedded in disks with relatively low 
${\dot M}_g$ are likely to capture each other onto their MMR's. 
This inference is consistent with the results in numerical model 
A1.  In this consideration, {\it the suppression of gas giant planet formation 
is due to the inability for embryos to merge rather than merely an 
inadequate supply of building block material.}


\subsection{Limited Gas Accretion}
\label{sec:gasacc}

Prior to the onset of efficient gas accretion, embryos with
$M_p < M_c$ can accrete gas, albeit on a Kelvin-Helmholtz 
cooling time scale $\tau_{\rm KH}$. For grain opacity with a
solar metallicity, $\tau_{\rm KH} \sim 10^{9} (M_p/M_\oplus)^{-3}$ yr 
\citep{Pollack-et-1996, Ida-Lin-2004a}.  In the limit that
$\tau_{\rm KH} > \tau_{\rm dep}$, embryos may accrete envelopes 
with mass $M_{\rm env} \sim M_p \tau_{\rm dep}/ \tau_{\rm KH}
\sim (M_p/M_\oplus)^4 (\tau_{\rm dep} / 1 \ {\rm Gyr}) M_\oplus$  
before the gas is depleted in the disk. Due to energetic impacts 
between embryos and residual planetesimals, this envelope mass may 
not be retained.  These diverse outcomes may contribute to the 
observed dispersion in the density of super Earths \citep{Wu-Lithwick-2013}.

The above consideration indicates that cores' $M_p$ need to 
exceed $M_c$ for them to evolve into gas giants before 
disk depletion.  Equation (\ref{eq:mopt}) indicates that 
$M_{\rm opt}$ is an increasing function of ${\dot M}_g$.  
For the disk parameters in model A1, $M_{\rm opt} < M_c$ 
such that the embryos' corotation torque is saturated 
before they evolve into cores.  We introduce model 
A2 to illustrate this inference. Four $10 M_\oplus$ embryos are
placed in a disk with identical parameters as those in model A1.
They are initially separated by $10 R_R$, starting from 0.43AU.
The results in the right panel of Figure \ref{fig:fig2} clearly
show that all embryos undergo rapid orbital decay. This model shows 
that {\it the successful assembly of super-critical-mass embryos 
does not guarantee their retention on a time scale comparable to 
either the gas accretion or disk depletion time scales.}

For the discussion of several competing processes, we 
adopt here steady state disk models.  They accentuate 
the potential of resumed migration for embryos with 
$M_p < M_c$ which failed to evolve into cores. In a 
subsequent paper, we will investigate embryos' orbital 
evolution as $r_{\rm trans}$ and $M_{\rm opt}$ decline 
with $\Sigma_g$ with ${\dot M}_g$ during the advanced 
stage of disk evolution.  

\section{Mergers and Super-critical Cores}
\label{sec:merger}
In this section, we first show that MMR capture may be 
bypassed in disks with sufficiently large ${\dot M}_g
(\gtrsim 10^{-7} M_\odot$ yr$^{-1}$).  They
converge into a compact region (with semi major axis 
separation $\Delta a < R_f$) where they cross each other's orbits.
High accretion rate
also obliterates two other growth barriers for embryos 
with overlapping orbits.  These obstacles are 1) large-angle 
scattering during close encounters, and 2) saturation of 
corotation torque and resumption of inward migration 
before cores are able to acquire a mass $M_c$. 

For the active disk simulations, we adopt, in model B, 
${\dot M}_g = 10^{-7} M_\odot$ yr$^{-1}$. For model B1, 
we place seven embryos 
between 5-15.3AU with $M_p = 5\rm M_{\oplus}$ and $4 M_\oplus$
for the inner two and outer five embryos respectively.
These values of $M_p$'s are comparable to $M_{\rm opt}$.
The embryos' semi major axes are initially separated by 
$10 R_R$. Models A1 and B1 have the same total mass.

\begin{figure*}[htbp]
\includegraphics[width=0.5\linewidth,clip=true]{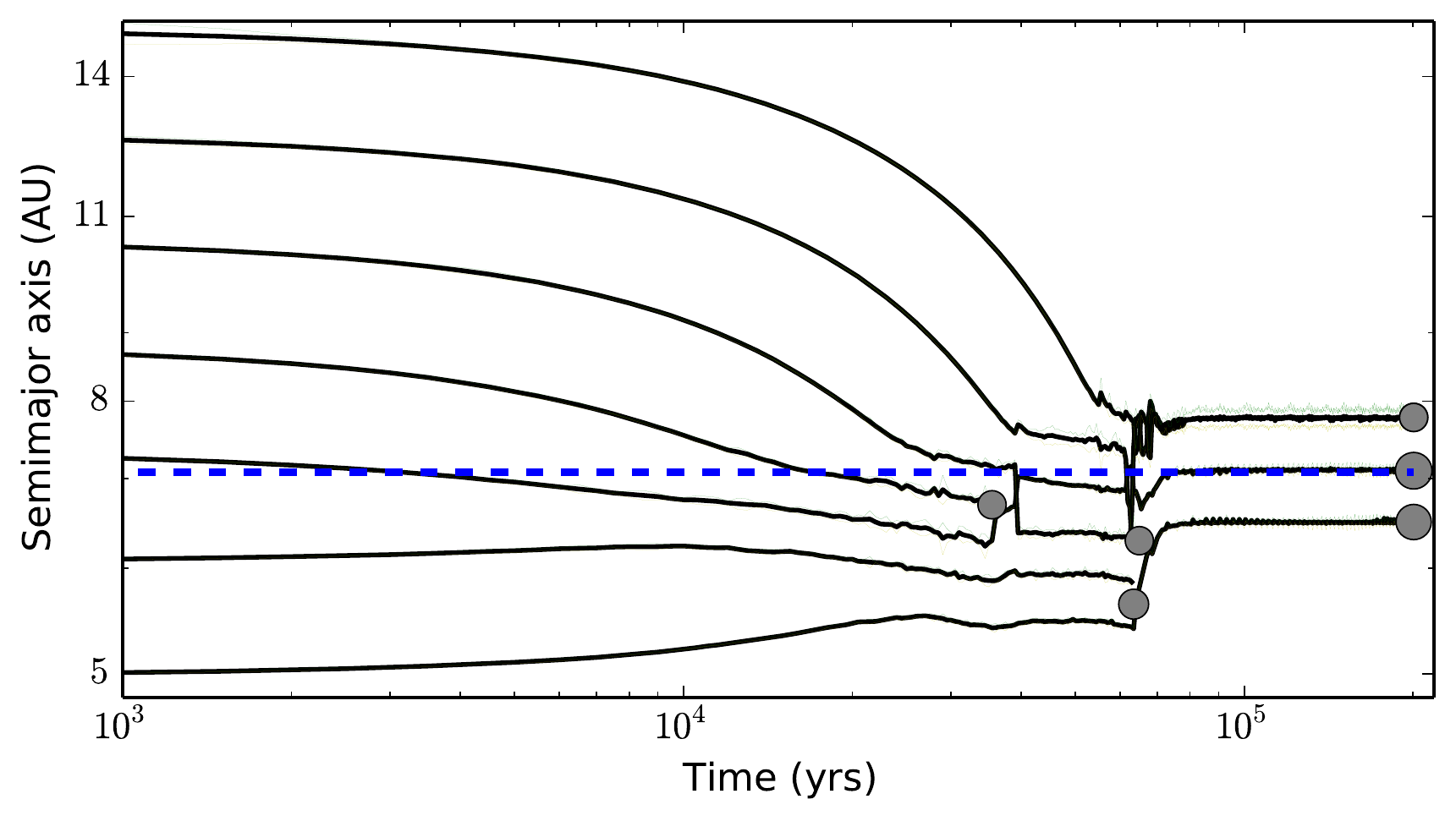}
\includegraphics[width=0.5\linewidth,clip=true]{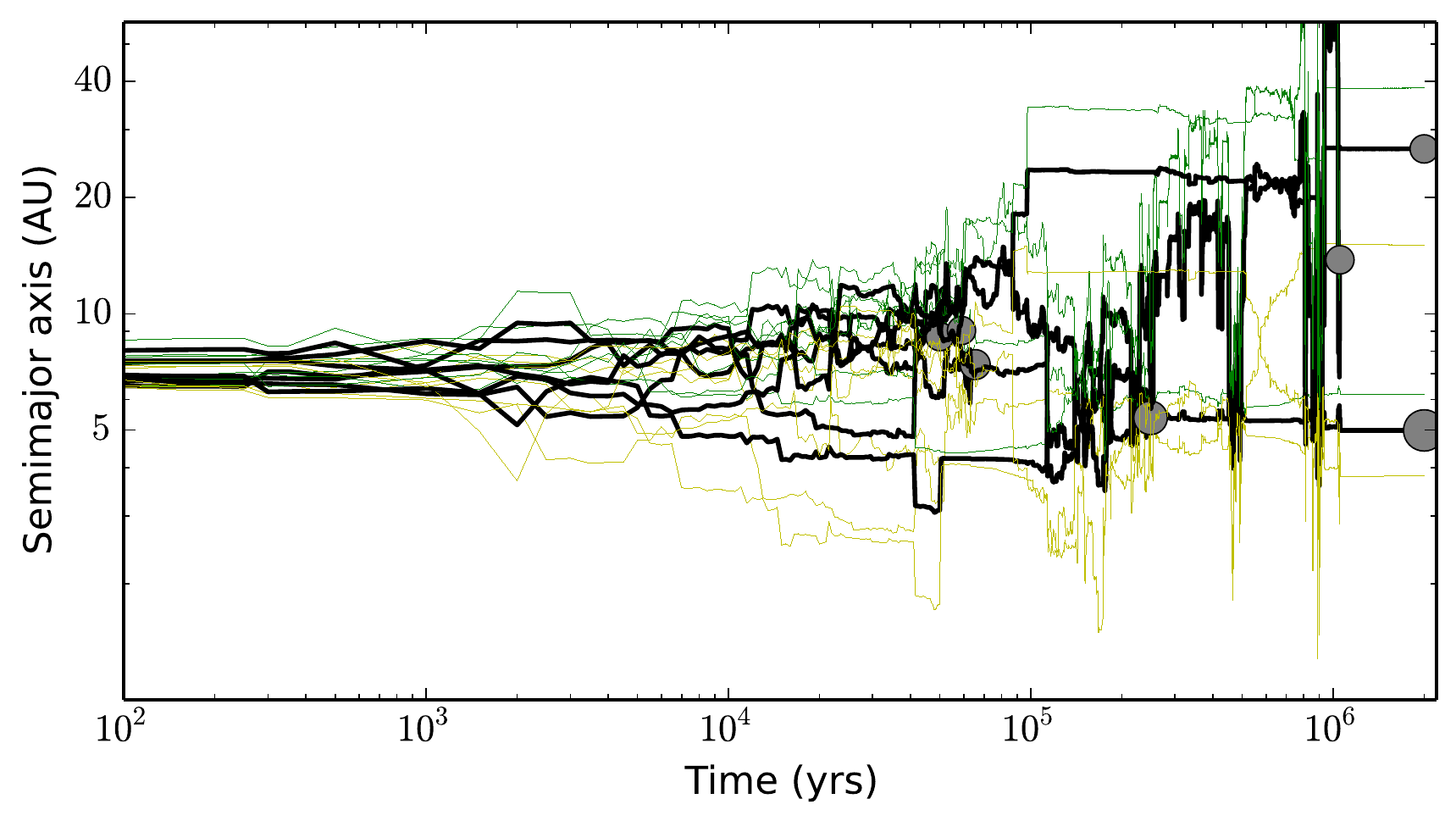}
\caption{
The mutual interaction between embryos and their natal disks. 
The black lines trace the evolution of embryos' semi-major axis and  
blue dashed line indicates the location of  $r_{\rm trap}$.  Green and 
yellow lines are embryos' apocentre and pericentre distance. 
 {\bf Left} model B1:  A system of seven embryos including inner two $5 M_{\rm \oplus}$  
and outer five $4 M_{\rm \oplus}$ embryos which are initially separated by 
$10 R_{\rm R}$ between $5.0-15.4$AU. Disk parameters are chosen to be
those of model B. 
{\bf Right} model B2:  A system of seven embryos including 
inner two $5 M_{\rm \oplus}$ and outer five $4 M_{\rm \oplus}$ 
embryos which are initially separated by 
$10 R_{\rm R}$ between $6.5-7.8$AU in the absence of a gaseous disk. 
}
\label{fig:fig3}
\end{figure*}

\subsection{Orbit Crossing and Close Encounters}
\label{sec:encounters}
Embryos migrate on a time scale $\tau_a \sim 2-3 \times 10^4$ 
yr and converge to $r_{\rm trans}$ as their separation is 
reduced to less than half of their initial spacing (see 
left panel of Figure \ref{fig:fig3}). The relatively
high value $\Sigma(r_{\rm trans}) >\Sigma(r_{\rm res})$ 
(see Eqs. [\ref{eq:sigmares}], [\ref{eq:sigmartrans}]  and   [\ref{eq:modtrtrans}] ) enables 
the embryos to avoid MMR capture.  

In such compact configurations, embryos' mutual perturbation 
excite each other's eccentricity with a growth time scale  
which decreases rapidly with their separation.  Using an 
idealized disk model in which the $\Sigma_g$ distribution
scales with the minimum mass nebula model, \citet{Zhou-et-2007}
estimate that the normalized width of the feeding zone is
\begin{equation}
{\rm log} k_{\rm iso} \simeq (b^2 + 0.61 c)^{1/2} -b 
\end{equation}
where $b = 2.8 + 0.33 {\rm log} \eta_Z$, $c = 3.6 + 0.67 {\rm log} 
\eta_Z + {\rm log} \tau_{\rm dep}$, $\eta_Z = \eta_d ^{3/2} 
(a/ 1{\rm AU})^{3/4} m_\ast^{-3/2}$, and $\eta_d$ is embryos' surface 
density enhancement factor (relative to the MMN model).  Embryos 
with $\Delta a < k_{\rm iso} R_R$ undergo orbit crossing within 
$\tau_{\rm dep}$.  For model B1, embryos'  $\Delta a$ is reduced 
to less than $5 R_R$ and their orbits begin to cross within a few 
$10^4$ yr.

After embryos enter each other's feeding zone, they undergo 
close encounters with impact parameter down to the embryos'
radius ($R_p$) which is $\sim 10^{-4} a$ 
and $<0.1$ times the size of typical computational 
mesh in hydrodynamic simulations.  The Hermite-Embryos scheme 
is designed and well suited to accurately integrate the 
orbital evolution associated with these close encounters.

Embryos with overlapping orbits undergo repeated close encounters
as they venture into each other's Roche radius.  Their eccentricity
is excited to $\sim R_R/a$ on a synodic time scale $\tau_{syn} 
\sim a P/R_R$.  It attains an equilibrium value $\sim (\tau_e/
\tau_{syn}) (R_R/a) \sim 0.02-0.03$ which is consistent with
the left panel of Figure \ref{fig:fig3}. Although the corresponding
Safronov number ($\Theta \sim 10^2$) significantly enlarges the 
embryos' cross section, they
scatter many times before any pairs physically collide.  In the 
proximity of $r_{\rm trans}$, some close encounters lead to large angle 
deflections, eccentricity excitation and semi major axis spreading.  
The strength of these perturbations is an increasing function of 
\begin{equation}
f_V \equiv V_e (M_{\rm opt}) / V_k (r_{\rm trans}).
\label{eq:fvdef}
\end{equation}
Equations (\ref{eq:mopt}) and (\ref{eq:trans1}) indicate 
respectively that $M_{\rm opt} \propto {\dot m}_9 ^{7/12}$ 
and $r_{\rm trans} \propto {\dot m}_9 ^{0.72}$ so that 
$f_V \propto {\dot m}_9 ^{0.5}$.  In model B1, close 
encounters with $f_V < 1$ weakly excite embryos' $e$'s.
Subsequently, {\it the disk torque not only damps the scattered 
embryos' eccentricity, but also repatriates them back to 
the proximity of $r_{\rm trans}$}.  

We highlight these effects with a comparative N-body simulation
(in which embryo-disk interaction is neglected).
In model B2, we place seven embryos with spatial order but much
closer separation (between 6.5-7.8 AU) than model B1.  The right
panel of Figure \ref{fig:fig3} show embryos' $e$ are excited
from negligible initial values to $\sim 0.5$.  As a consequence 
of the close encounters their semi major axes $a$ also become 
widely separated.  Comparison between models B1 and B2 indicates 
that in disks with sufficiently large $\Sigma_g$ (and ${\dot M}_g$), 
embryos remain congregated near $r_{\rm trans}$ because the 
embryo-disk torque is effective to damp embryos' eccentricities 
$e$ and to repatriate them back to 
$r_{\rm trans}$ for repeated encounters.

In disks with sufficiently high accretion rate 
(${\dot M}_g > 2 \times10^{-7} M_\odot$ yr$^{-1}$),
$r_{\rm trans}$ exceeds $\sim 10$ AU where $f_V \gtrsim 1$ 
for embryos with $M_{\rm opt}$. Many embryos are 
episodically scattered into highly elliptical orbits.  
Even though embryos resume their convergent migration
and orbital circularization, this effect significantly
prolongs the time scale for embryos to grow through
cohesive collisions.  

\subsection{Embryos' Collisions}
Eventually, the N embryos within  
$\Delta a$, collide and merge on a time scale
\begin{equation}
\tau_c  \sim (a \Delta a / N R_p^2) P /2 \pi \Theta
\label{eq:tauc}
\end{equation} 
where $\Theta = G M_p/R_p \sigma^2$ is the Safronov number
and $\sigma \sim \Delta a V_k/a$ is the velocity dispersion.
For a convoy of a few super Earths with $\Delta a \sim R_R$, 
$\tau_c \sim 10^4 P$ at 
$a \sim $ several AU's.  Although $\tau_c$ is short compared with the 
disk lifetime (a few Myr), 
it is sufficiently long to render 3D hydrodynamic simulation of the 
merging process impractical. This technical issue is particularly
acute for disks with large ${\dot M}_g$ where $r_{\rm trap}$ is 
at a few AU's and trapped embryos' close encounters can significantly 
enlarge their $\Delta a$ and prolong their $\tau_c$.

This computational challenge may be partially reduced with 2D simulations
of Z14a, Z14b
in which $\tau_c$ is shortened by 
a factor of
  $f_{2/3D} \simeq 1 + \langle i \rangle a/R_p$
where $\langle i \rangle$ is the 
average inclination of the embryos.  
We note the enhancement factor 
would be $   a \langle e \rangle /R_p$ if the velocity dispersion is isotropic ($\langle i \rangle  \sim  \langle e \rangle$) and  would be
unity if the system is a mono-layer ($\langle i \rangle  \sim  0$).
However, even in the mono-layer  
limit, it is impractical to adequately explore the model
parameter space and determine the destiny of embryos during the disk 
evolution.

In our simulations, embryos with overlapping orbits repeatedly undergo close 
encounters until they physically collide with each other.
The first pair of embryos cross each other's orbits at $\sim 7$ 
AU (near $r_{\rm trans}$) after $\sim 3.5 \times 10^4$ yr.
They then collided with each other within a few hundred periods
which is consistent with both the results in Z14b and our 
estimate of $f_{2/3D} \tau_c$ in equation (\ref{eq:tauc}). 
The magnitude of $\tau_c$ is $f_{2/3D} \sim 10^2$ longer 
in the 3D limit \citep{Rafikov-2004}.  However, if the embryos'
inclination distribution is damped to that of a mono layer, 
the 2D estimate would be appropriate.
Since both  3D linear analyses \citep{Tanaka-et-2002} and  full 3D hydrodynamic simulations \citep{Bitsch-Kley-2011} indicate the inclination damping timescale is nearly the same order as
eccentricity damping timescale, the above assumption is well justified.

Two additional mergers occurred within $7 \times 10^{4}$ yr.  
The mass of these merger products became comparable to the 
critical core mass $M_c$. Two outermost cores captured each other, 
remained locked in a co-orbital resonance at $7.7$AU 
(left panel of Fig. \ref{fig:fig3}) within $2 \times 10^{5}$ yrs 
and may eventually merge.  The mass of the merger products 
approaches to the critical value ($M_c$) for the onset of 
efficient gas accretion at around $7.1$ AU and $6.5$ AU.  

For embryos with $M_p \sim M_c (\sim 10 M_\oplus)$, 
$\tau_{\rm KH} < \tau_{\rm dep} (\sim$ 3-5 Myr). The 
magnitude of $\tau_{\rm KH}$ may be reduced due to an opacity 
reduction (from its values in the the interstellar medium 
with solar composition) associated with grain sedimentation 
in the protoplanetary envelope \citep{Ikoma-et-2000,
Helled-Bodenheimer-2011}.  
Provided the cores' $M_c$ does not substantially exceed 
$M_{\rm opt}$, they are retained near $r_{\rm trans}$ 
before they gain sufficient mass to open a gap near their 
$a$'s.  In model B1, the magnitude of $M_{\rm opt}$ is 
much larger than that in model A2.  In fact, $M_{\rm opt} 
\sim M_c$ which implies that cores, once assembled, are more 
likely to be retained in disks with high ${\dot M}_g$'s.

\subsection{Trapped Embryos' Mass Range}

Figure~\ref{fig:fig1} indicates that the corotation torque 
is saturated for small embryos with $M_p < 1 M_\oplus$ for 
model A and $M_p < 3.5 M_\oplus$ for model B.  With a
uniform $\alpha$ prescription, we carried out simulations 
with 15 low-mass ($2 M_\oplus$) embryos (model B) and 
confirm that they indeed migrate inward, albeit at modest 
speeds because $\Gamma_0$ is relatively smaller for low-mass 
planets (see eq. [\ref{eq:gamma0}]).  

In \S\ref{sec:disk} and \S\ref{sec:migdisk}, we indicate that 
the existence of a dead zone with active surface layers modifies 
the saturation of the corotation torque. We construct model C 
with a set of identical disk structure parameters as those in
model B.  But in the calculation of $f_a$, we used a prescription
(similar to KL12) in which 
$ \alpha_\nu = \alpha_{ M} $  when $ \ R_{R} <  R_{\rm dz}$ and 
 $ \alpha_\nu=\alpha_{\rm M}+(\alpha_{H} -\alpha_{M}) \left( \frac{(R_{\rm R}/R_{\rm dz})^{2}-1 } {(R_{\rm R}/R_{\rm dz})^{2}+1}\right)$  
when $R_{H} > R_{\rm dz}$. $R_{\rm R}$ is the planetary Roche radius
  and the size of dead zone $R_{\rm dz}= H(\rm r) \Sigma_{g}(\rm r)/\Sigma_{\eta} $, 
  where  $\alpha_{ H}=10^{-3}$, $\alpha_{ M}=1.4 \times 10^{-4}$  and   $\Sigma_{\eta} $ is a scale value  independent  of $\rm r$.

\begin{figure}[htbp]
\includegraphics[width=0.98\linewidth,clip=true]{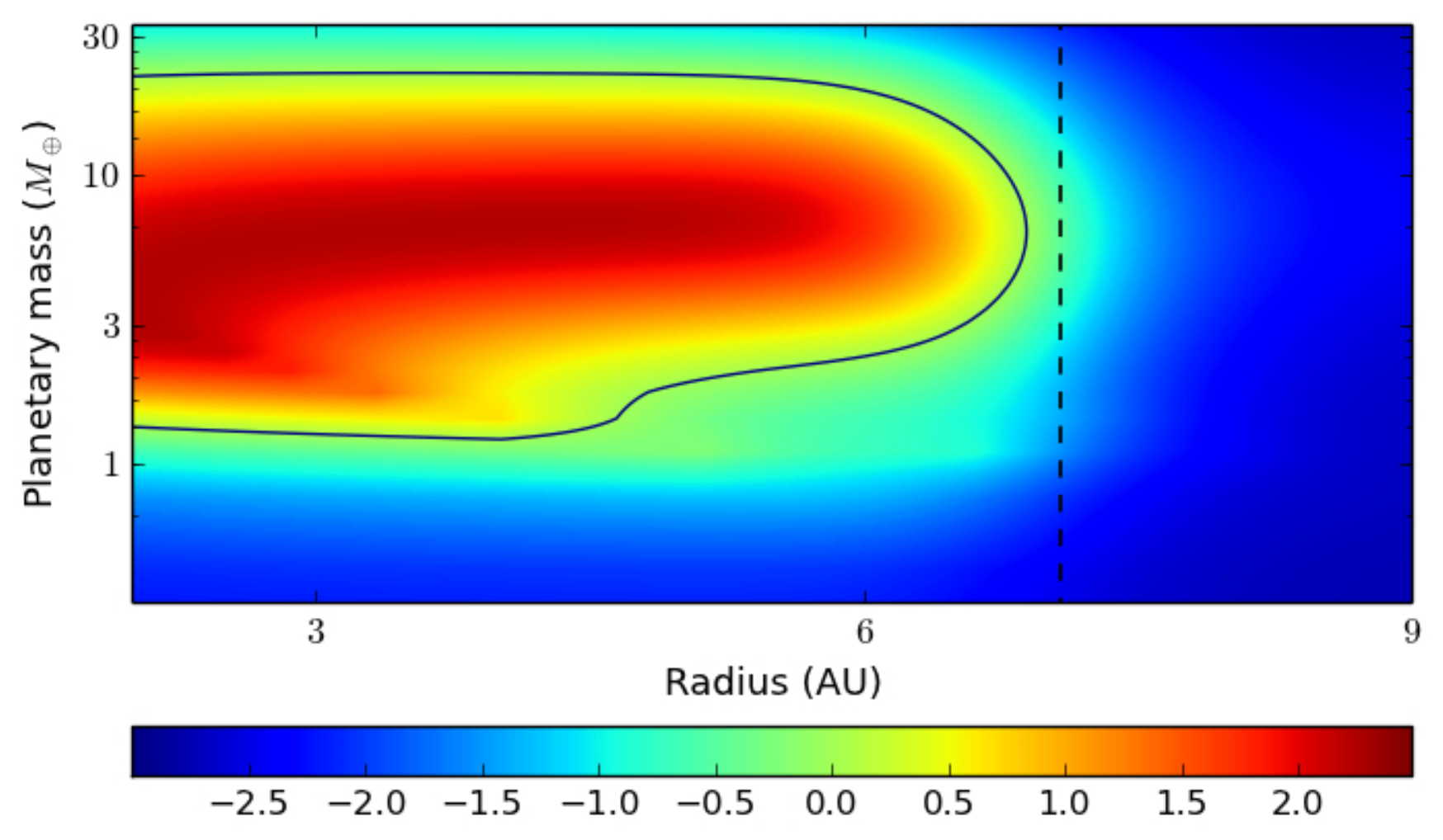}
\includegraphics[width=0.98\linewidth,clip=true]{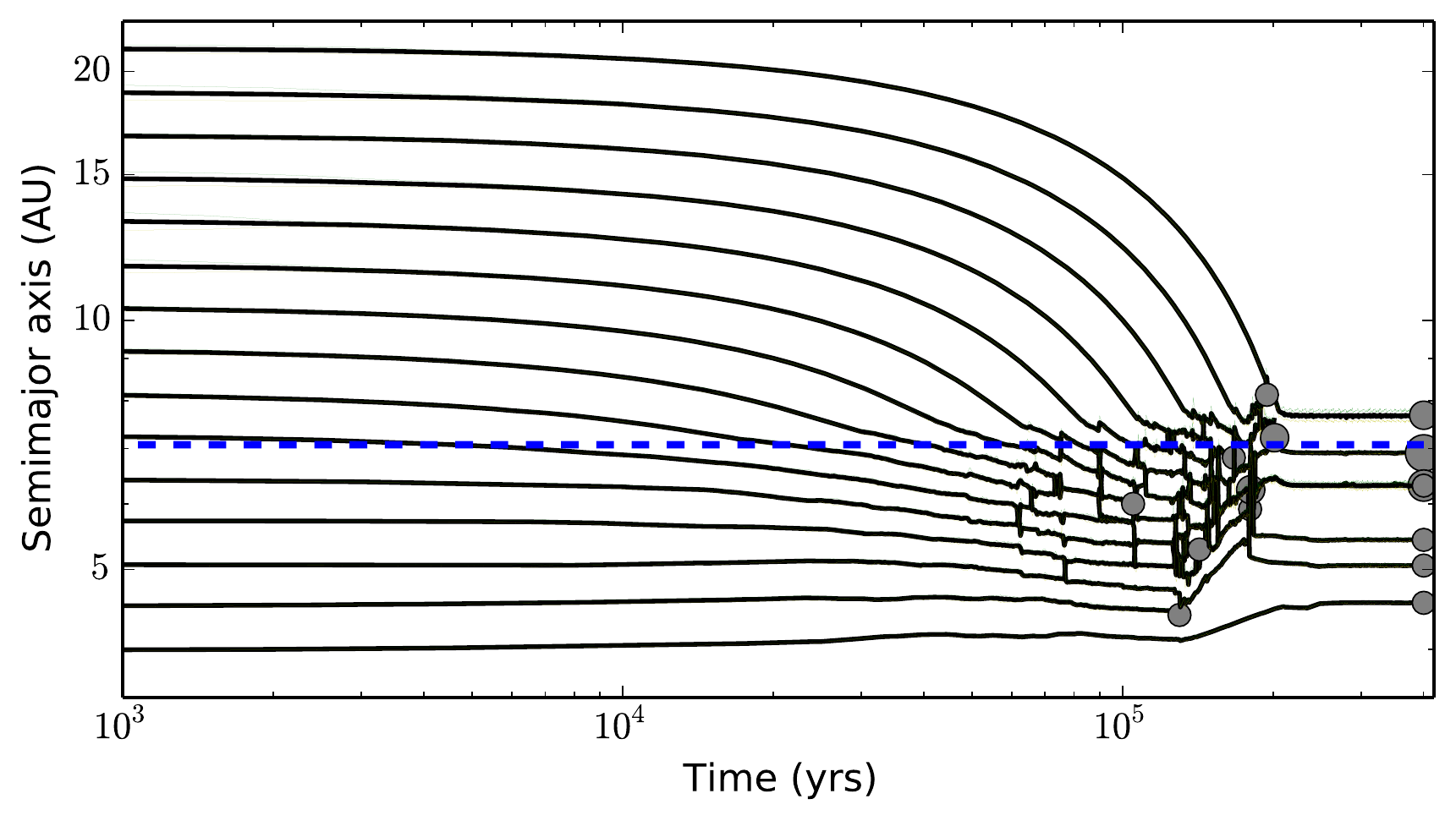}
\caption{
{\bf Top}: 
The type \uppercase\expandafter{\romannumeral1} migration coefficient  ($f_a$) 
varies with different location and planetary mass at the onset of the
simulation for Model C.  In this model, the layered structure for the 
dead zone in the disk is taken into account with a prescription.  The 
disk parameters are chosen the same as those in model B in the right 
panel of Fig.~\ref{fig:fig1}.
{\bf Bottom}: Model C1: the dynamical evolution of multiple embryos 
embedded in  above disk model. Fifteen embryos with $2 M_{\oplus}$ 
are initially separated by $8R_{\rm H}$ from $4$AU to $21.3$AU .
}
\label{fig:fig4}
\end{figure}

The top panel of Figure \ref{fig:fig4} indicates that this
prescription does not modify $r_{\rm trans}$ but it does 
enlarge the mass range ($1.5 M_p \sim 25 \rm M_{\oplus}$) for the 
outwardly migrating embryos.
In model C1, we place 15 embryos, each with a mass $M_p=2
M_\oplus$ (as in model A1), initially separated by $8 R_R$ between
4 and 21.3 AU.  In contrast to model B2, embryos initially located 
 at $a<r_{\rm trans}$ migrate outward.  They converge with the 
inwardly migrating embryos on to confined regions with overlapping
orbits.  Similar to the results in model B1, the first collision
(at $\sim  10^5$ yr) was followed by several others.  
Within $ 2 - 3 \times10^5$ yr, seven embryos remain and maximum embryo mass  $M_p$ attains 
$12 M_\oplus$ (located at 6.9 AU). In model C1, embryos  more massive 
than $M_c$  can be  retained near $r_{\rm trans}$ in contrast 
to the results in model A2.  

\section{Summary and Discussions}
\label{conclusions}

Gas giant planets are found around $ 15-20 \%$ of nearby 
solar type stars. In the sequential accretion scenario, they
are formed through gas accretion onto protostellar cores.  The
accretion rate is determined by the efficiency of radiation 
transfer through the gaseous envelope. Its associated Kelvin
Helmholtz contraction time is a steeply decreasing function of
the cores' mass.  Cores can grow into gas
giants prior to severe disk depletion only if their mass
exceeds a critical value of 10 $M_\oplus$.  

The assemblage of critical mass ($M_c$) cores is a crucial 
step in the formation of gas giant planets. These objects formed 
through the coagulation of smaller protoplanetary embryos 
whose oligarchic growth is quenched when they consume all 
the building block planetesimals in their feeding zone.  
Embryos' dynamical isolation mass at a few AU in a MMN 
is a few $M_\oplus$ \citep{Ida-Lin-2004a}.

In this paper, we adopt the assumption that
migration plays a significant role in dynamical architecture 
and final fate of planetary systems
\citep{Nelson-2005,Alibert-2005,Masset-et-2006,Paardekooper-et-2011}.
We present simulations here to show that one possible
mechanism to enlarge the isolation mass is through embryos' 
extensive type I migration \citep{Lyra-et-2010, Horn-et-2012, Hellary-Nelson-2012,Pierens-et-2013}.  
We constructed a Hermite-Embryo code which
includes embryos' interaction with their natal disk and with
each other.  For the embryo-disk torque, we applied  
existing prescriptions into a self consistent disk model.

For the viscously heated inner regions, we show that at its 
full strength, corotation torque 1) transfers angular momentum 
from the disk to the embryos at a rate faster than that due to 
the differential Lindblad torque and 2) induces embryos to 
migrate outward.  However, the corotation torque is saturated
(i.e. suppressed) for both relatively high and low mass embryos.  
For the outer region which is heated by stellar irradiation, 
embryos generally migrate inward.  These embryos converge at
the interface between these regions (typically at a few AU's).

Our results indicate that in disks with ${\dot M}_g <10^{-7} 
M_\odot$ yr$^{-1}$, embryos are caught in their mutual MMR as 
they slowly approach each other. In this limit, they cannot coagulate 
and attain the critical mass needed to evolve into gas giants,
though they may still accrete a modest envelope. 
Many of these super Earths are found in 
multiple systems with a total mass in excess of 10 $M_\oplus$. 
We suggest they are the embryos which failed to attain $M_c$ and 
evolve into cores. The results in Figure \ref{fig:fig6} indicate that
the minimum total available building block materials around the host 
stars of most multiple systems are more than adequate to form super 
critical mass cores. But most of them do not bear signs of gas giant 
planets. We interpret these data to imply that the lack of gas giants 
around most solar type stars may be due to the inability 
for sufficient fraction of all available building block materials 
to be collected into a few super-critical cores (with $M_p \geq M_c$)
rather than a limit supply of heavy elements in their natal disks
\citep{Laughlin-2004,Ida-Lin-2004b,Mordasini-et-2009}.

We also show here that embryos' convergent speed increases with the gas accretion rate.
In disks with ${\dot M}_g \gtrsim 10^{-7} M_\odot$ yr$^{-1}$, 
embryos congregate with overlapping orbits around a trapping 
radius outside 7AU.  They undergo repeated close encounters 
while the disk torque damps their excited eccentricity and 
repatriates them back to the trapping location. The concentration 
of embryos elevates their isolation mass and leads to 
the assemblage of cores. 

The threshold criteria (${\dot M}_d \sim 10^{-7} M_\odot$ 
yr$^{-1}$) is estimated for a steady disk with an assumed 
$\alpha_\nu \sim 10^{-3}$.  This value is consistent with
numerical simulations of MRI disks  \citep{Sano-et-2004,
Fromang-Nelson-2006} and that
infered from modeling the observed disk accretion rates and 
masses.  We also note that $\sim 20\%$ protostellar disks 
around T Tauri stars have ${\dot M}_g \gtrsim 10^{-7} M_\odot$
yr$^{-1}$. This distribution function provides some support
for our conjecture that gas giants around solar type stars 
are preferentially formed in high-${\dot M}_g$ disks.  All 
of these estimates are somewhat uncertain.

Different mechanisms have been attributed as the dominant cause of  
"planet trap" \citep{Masset-et-2006}, including the separatrix of turbulent inner region and 
outer dead zone \citep{Morbidelli-et-2008}, transition of opacity 
\citep{Lyra-et-2010,Bitsch-et-2013}, or transition of dominant energy budget 
discussed by \cite{Kretke-Lin-2012} and this paper. Albeit with
some discrepancies,  the outward migration mass range and 
transition radius are universally shown as well. The disk
structure promotes the embryos' convergent migration and the 
accumulating them near different proposed  $r_{\rm trap}$.

Disk models with a broad range of structural parameters
including the total disk mass, radial and vertical distribution of
viscosity, opacity, accretion rate, detailed energy budget have been 
applied to hydrodynamical simulations of planet-disk tidal interaction
\citep{Bitsch-et-2013,Bitsch-et-2014} and N-body plus additional 
analytical force simulations \citep{Hellary-Nelson-2012, Pierens-et-2013} similar to our simulation with the Hermite-Embryo code. 
Despite the diversity in these disk models,  the simulations 
nevertheless confirm the robustness of embryos' convergent migration process and 
indicate that embryos' migration history determines whether they evolve
into super Earths or cores of gas giant planet.

\cite{Hellary-Nelson-2012} simulated the embryos' convergent migration and  growth in non-isothermal but  somewhat arbitrary chosen disk profile.   
  \cite{Pierens-et-2013}  suggested that the resonant convoy 
can be broken with a large initial number of embryos (total planetary mass in disks) or by including 
a moderate stochastic force due to the disk turbulence. 
Although \cite{Hellary-Nelson-2012} and  
 \cite{Pierens-et-2013} also mentioned briefly how the disk mass may affect the location of 
$r_{\rm trap}$, they did not discuss its influence on the 
embryos' ability to bypass the MMR.  
After the submission of this paper, \citet{Cossou-et-2014} 
posted on Arxiv.org results obtained with a similar approach
but a different disk model.  In contrast to the self-consistent
steady state disk model (GL07), they assumed a surface density profile 
and derive a temperature distribution by combining contributions from
viscous heating, stellar irradiation and radiative cooling. In their 
evolving-disk model, opacity and $\dot M$ vary with radius and $r_{\rm trap}$
is located near the opacity transition region where the disk temperature 
gradient is steep.  They obtained similar results but did not obtain
the quantitative criteria presented here.  Another recent paper
by \citet{Coleman-Nelson-2014} simulated both the formation of cores, gas
accretion, and the gas giants' type II migration in evolving disks.
Based on their simulation results, they
suggested that gas giants formed at large radii in a sufficiently 
late epoch are preferentially retained.  However, many gas 
giants are observed to reside in multiple-planet systems.  Their formation
requires adequate residual gas and embryos in their natal disks.
The results in Figure \ref{fig:fig2} (Model A2) indicate that 
the corotation torque between super Earth cores and low-
${\dot M}_g$ disk is also quenched by saturation.  Unless they 
can induce a gap and a transition to type II migration, these
cores would not 
be retained when the disk gas is severely depleted.

Inspired from the observation (See Figure \ref{fig:fig6} and the  difference between $\eta_J$ and  $\eta_\oplus$), our theoretical analysis and numerical simulations places a strong 
emphasis on that ubiquitous presence of super Earths and limited 
frequency of gas giants around solar-type stars are the manifestation 
of a threshold condition which depends on the magnitude of disk 
accretion rate ${\dot M}_g $.
The results indicate that the embryos must undergo relatively fast 
convergent migration in order to bypass the MMR barriers and merge 
into super-critical cores with $M_p> M_c \simeq 10 M_{\oplus}$.
Although previous investigations produced similar results for the 
condition of multiple embryos to overcome resonant barriers, including
the dependence of $r_{\rm trap}$ on the disk mass
 \citep{Hellary-Nelson-2012,Pierens-et-2013,Cossou-et-2014}, they did not 
discuss the dependence of the merger probability on the disk mass
and accretion rate.


\vskip 20pt

\acknowledgments
We thank Drs C. Baruteau, S. Ida, K. Kretke, H. Li, 
and T. Kouwenhoven for useful conversations and an anonymous referee
for helpful suggestions to improve the presentation.  This 
work is supported by grants from  LDRD, IGPPS from LANL, and UC/Lab Fee's 
program.  B. Liu also thanks T. Kouwenhoven for support with an NSFC grant.  

\bibliographystyle{apj}
\bibliography{reference}

\end{CJK}
\end{document}